\documentclass[
aps,%
12pt,%
final,%
notitlepage,%
oneside,%
onecolumn,%
nobibnotes,%
nofootinbib,%
superscriptaddress,%
noshowpacs,%
centertags,%
natbib
]{revtex4} 

\usepackage{epstopdf}
\usepackage[usenames]{color}
\usepackage{graphicx,times}
\usepackage{natbib}
\usepackage{amssymb,amsmath}

\usepackage{xcolor}
\usepackage{lscape}
\usepackage{longtable}
\usepackage{array}
\usepackage{lscape}
\usepackage{xtab}

\begin{document}

\title{Relation between parameters of dust and parameters of molecular and atomic
gas in extragalactic star-forming regions}

\author{\firstname{K.~I.}~\surname{Smirnova}}
\email[E-mail: ]{Arashu@rambler.ru}
\affiliation{Ural Federal University, Ekaterinburg, Russia}

\author{\firstname{M.~S.}~\surname{Murga}}
\affiliation{Institute of Astronomy, Russian Academy of Sciences, 
Moscow, Russia}

\author{\firstname{D.~S.}~\surname{Wiebe}}
\affiliation{Institute of Astronomy, Russian Academy of Sciences,
Moscow, Russia}

\author{\firstname{A.~M.}~\surname{Sobolev}}
\affiliation{Ural Federal University, Ekaterinburg, Russia}

\begin{abstract}
The relationships between atomic and molecular hydrogen and dust of various sizes in extragalactic star-forming regions are considered, based on observational data from the Spitzer and Herschel infrared space telescopes, the Very Large Array (atomic hydrogen emission) and IRAM (CO emission). The source sample consists of approximately 300 star-forming regions in 11 nearby galaxies. Aperture photometry has been applied to measure the fluxes in eight infrared bands (3.6, 4.5, 5.8, 8, 24, 70, 100, and 160$\mu$m), the atomic hydrogen (21~cm) line and CO (2--1) lines.

The parameters of the dust in the starforming regions were determined via synthetic-spectra fitting, such as the total dust mass, the fraction of polycyclic aromatic hydrocarbons (PAHs), etc. Comparison of the observed fluxes with the measured parameters shows that the relationships between atomic hydrogen, molecular hydrogen, and dust are different in low- and high-metallicity regions. Low-metallicity regions contain more atomic gas, but less molecular gas and dust, including PAHs. The mass of dust constitutes about ~$1\%$ of the mass of molecular gas in all regions considered. Fluxes produced by atomic and molecular gas do not correlate with the parameters of the stellar radiation, whereas the dust fluxes grow with increasing mean intensity of stellar radiation and the fraction of enhanced stellar radiation. The ratio of the fluxes at 8 and 24~$\mu$m, which characterizes the PAH content, decreases with increasing intensity of the stellar radiation, possibly indicating evolutionary variations of the PAH content. The results confirm that the contribution of the 24~$\mu$m emission to the total IR luminosity of extragalactic star-forming regions does not depend on the metallicity.

\end{abstract}

\maketitle

\section{INTRODUCTION}

Star formation is a complex set of phenomena that unites many different physical processes. Star formation occurs in regions with high densities of interstellar matter, which often host clusters of massive young stars that ionize the surrounding gas. Such objects are often called ionized hydrogen (HII) regions or star-forming regions (SFRs). Star formation in spiral galaxies mostly occurs in the spiral arms. The spatial distribution of SFRs in irregular galaxies can be chaotic. The chemical composition of the medium significantly affects the behavior of star formation, since it determines, e.g., the gas cooling and heating rates.

The main components of the interstellar medium (ISM) that take part in star formation are atomic
and molecular gas, and also dust. The presence of atomic gas is traced by emission in the 21-cm neutral hydrogen line. Molecular hydrogen essentially does not emit under the conditions typical for SFRs, and information about the distribution of molecular gas is primarily extracted from observations of other molecules, most importantly carbon monoxide (CO). Continuum observations at wavelengths between several microns and one millimeter and observations of near infrared (NIR) emission bands are a source of information about various dust components. Modern ground- and space-based instruments can
be used to observe and spatially resolve such indicators in extragalactic SFRs, making it possible
to search for relations between different components of the ISM and star  formation properties  over  wider ranges of parameters (e.g., metallicities) than in our Galaxy

The interrelationships between different ISM components have been explored in numerous studies,
which are being developed in several directions. Comparisons between tracers of atomic and molecular
gas and indicators of the star-formation rate (such as different emission lines or IR emission from
dust) enable studies of the dependence of the  star formation rate  on  the  surface  density  of  gas  or  the  degree  of of molecularization of the gas, etc. (see, e.g., \cite{Leroyetal2008,Wuetal2015}). Comparisons of data on the emission of molecular gas and dust are used to estimate possible variations of the so-called ``X-factor'' \cite{sandstrometal2013}, which is applied to convert the integrated intensity of CO lines into an H$_2$ column density.

\begin{table}[h!]
\caption{Data on the explored galaxies}
\label{galdata}
\begin{center}
\begin{tabular}{|l|c|c|c|c|c|c|}
\hline
\multicolumn{1}{c|}{Galaxy} &
\multicolumn{1}{c|}{$\alpha$} & \multicolumn{1}{c|}{$\delta$} &
\parbox[c][3cm]{2cm}{Distance, Mpc}&
\multicolumn{1}{c|}{\parbox[c][1cm]{2cm}{Morpho- logical type}}&
12 + log(O/H) &
\parbox[c][1cm]{2cm}{Number of regions}\\
\hline
NGC 628  & 24.1739458 & 15.7836619 & 9.0 \cite{Rodriguez2014} & SA(s)c & 8.45  &  56 (65) \\
NGC 925  & 36.8203333 & 33.5791667 & 9.0 \cite{Tully2013} & SAB(s)d & 8.34  &  30 (43) \\
Ho II    & 124.770750 & 70.720028 & 3.3 \cite{Tully2013} & Im & 7.72  &  \phantom{0}7 (12) \\
NGC 2976 & 146.814417 & 67.916389 & 3.6 \cite{Tully2009} & SAc pec & 8.36  &  5 (7) \\
IC 2574  & 157.097833 & 68.412139 & 3.8 \cite{Tully2009} & SAB(s)m & 7.85  &  11 (16) \\
NGC 3351 & 160.990417 & 11.703806 & 10.1 \cite{Russell2002}\phantom{0} & SB(r)b  & 8.71  &  13 (23) \\
NGC 3627 & 170.0623508 & 12.9915378 & 9.9 \cite{Courtois2012} & SAB(s)b & 8.34 & 11 (13)  \\
NGC 4736 & 192.721088 & 41.120458 & 5.0 \cite{Tully2009} & (R)SA(r)ab & 8.44  &  10 (14) \\
DDO 154  & 193.521875 & 27.149639 & 4.1 \cite{Jacobs2009} & IB(s)m & 7.54  &  2 (2)\\
NGC 5055 & 198.955542 & 42.029278 & 8.2 \cite{Sorce2014} & SA(rs)bc & 8.65  &  23 (28) \\
NGC 6946 & 308.7180150 & 60.1539150 & 5.6 \cite{Rodriguez2014} & SAB(rs)cd & 8.47 & 46 (54) \\
\hline
\end{tabular}
\end{center}
\end{table}

Comparison of the properties of the 8~$\mu$m emission and of the CO emission presented in \cite{2010MNRAS.402.1409B} showed that, on large scales, the emission of molecular gas is correlated with the emission of small dust grains, namely, polycyclic aromatic hydrocarbons (PAHs; in particular, their radial profiles have statistically equivalent scale lengths); however, this correlation weakens on smaller scales. Such relationships are interesting because they shed light on the origin of so-called ``CO-dark'' molecular gas, i.e., gas with a very low CO abundance, even though  most of the hydrogen has already been converted into molecular form. In this case, molecular clumps could be traced using dust emission.
However, another conversion factor must be applied gas-to-dust ratio, which enables estimation of the amount of molecular hydrogen from the intensities of the thermal emission of various types of dust grains(see, e.g.,\cite{2012ApJ...761..168E,2014A&A...563A..31R,Groves}). The importance of atomic gas for star-formation processes should also be borne in mind. A number of studies describing the relationship between SFRs in HI-dominated galaxies and their metallicities \cite{SambitRoychowdhury2015}, as well as the properties of dust emission, have been carried out.

In our present study, we con sider the relationships between various gas and dust components of the ISM from  the  viewpoint  of  dust  evolution  rather  than  of star-formation  processes.
The  global  evolution  of dust in galaxies is currently considered to be largely determined  by  processes  taking  place  in  molecular clouds in SFRs.   Determining the interrelationships
between the properties of various dust populations in SFRs  and  in  the  ISM  as  a  whole  may  provide  an important contribution to studies of these processes. It is believed that the diverse properties of cosmic dust grains can described approximately by the following main  populations:   large  grains  with  radii ${>}100$~\AA), very small grains (VSGs) with radii several dozens of angstrom units,  and PAHs with radii of about  10~\AA) \cite{DustEM}. According to current views, both  silicate  and carbonaceous  large  dust  grains,  are  fairly  cool  and emit  mainly  in  the  far  infrared  (FIR),  submillimeter, and millimeter continuum (at wavelengths longer than 70~$\mu$m). 
Stochastic heating of VSGs (predominantly carbonaceous) provides emission in the middle
infrared  (MIR, about 20--30~$\mu$m). Finally, emission bands in the NIR range (from 3 to 20~$\mu$m) are  ascribed  to small  carbonaceous  dust  grains  that  contain  about 1000  or  fewer  C  atoms  and  have  high  abundances of aromatic compounds. These  grains  are  usually
called  PAHs,  although  their  structure  is  probably more complex \cite{jones2012_1}. Given the available observational data, PAH emission in extragalactic SFRs is usually studied  using  8-$\mu$m Spitzer IRAC data, and VSG emission can be studied using the 24~$\mu$m band of the MIPS spectrometer of the same telescope. Emission of larger grains at wavelengths longer than 70~$\mu$m can be detected with  the Herschel PACS and SPIRE instruments, as well as ground-based millimeter and submillimeter telescopes.

Such studies have been performed earlier either for entire galaxies or for individual ``pixels'' spatially resolving galaxy images. In particular, a comparison of dust emission at 8 and 24~$\mu$m showed that PAHs aremainly associated with grains in diffusemedia that emit at 160~$\mu$m~\cite{Bendoetal2008}, it was concluded that the ratio of the flux densities at 8~$\mu$m and 160~$\mu$m is an indicator of the intensity of interstellar radiation heating diffuse dust in galaxies. Another interesting conclusion is that PAHs can be used as a probe of the distribution of diffuse dust in galaxies, and that PAHs cannot be used (or must be used with caution) as indicators of star formation.

As we are interested in various aspects of the evolution of dust particles, we preferred to consider individual SFRs, where we expect the most prominent and rapid variations of dust-grain properties, rather than the galactic ISM in general. Our study was motivated by the work of Khramtsova et al. ~\cite{pahcycle}, who showed that the time variations of the relative abundances of dust grains of various types in extragalactic SFRs depend on the SFR metallicities.

As the resolution of the images we have used is fairly low ($12^{\prime\prime}$, see below), the linear scales of the studied objects are about 500 pc. Although this corresponds to the largest regions of coherent star formation, in most cases, we have determined the parameters of groups of closely spaced SFRs with characteristic sizes of about 100 pc that are unresolved in the IR images, rather than the parameters of star-forming complexes (although some of the regions studied could be real star-forming complexes). Therefore, our regions may contain several SFRs with different characteristics (age, mass, etc.), so that we have observed the net emission of all SFRs within the aperture.

Since we are not able to take into account differences between the ages of the objects, the list of studied SFRs may contain both SFRs in which star formation has recently started and those whose star formation is nearing completion. Note that the range of ages of SFRs considered in our previous studies \cite{pahcycle,HolmbergII}, which were selected in a similar manner, proved to be moderate, from three to eight million years.

Rather than the SFR ages, we have focused on their metallicities; it is unlikely that the metallicity varies strongly within a group of closely spaced regions, since the spatial scales for metallicity variations are much larger. The real internal structures of the objects may be fairly complex, and include cavities created by supernova explosions or stellar wind, filaments, and dense clumps mixed with tenuous medium. However, our aim was to determine the relationships between the integrated characteristics of gas and dust, independent of how they are distributed inside the objects.

\begin{figure}
\includegraphics[width = 7cm, angle=0]{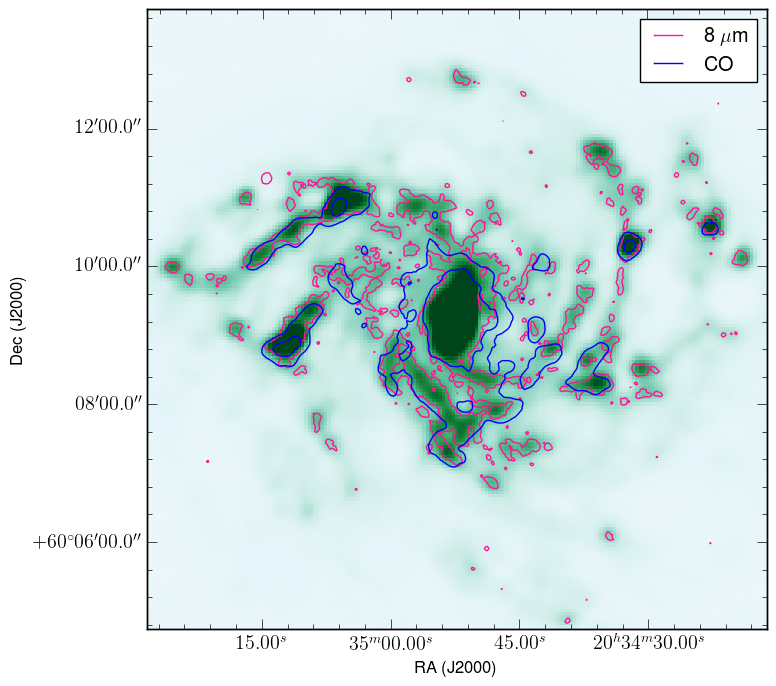}
\includegraphics[width = 7cm, angle=0]{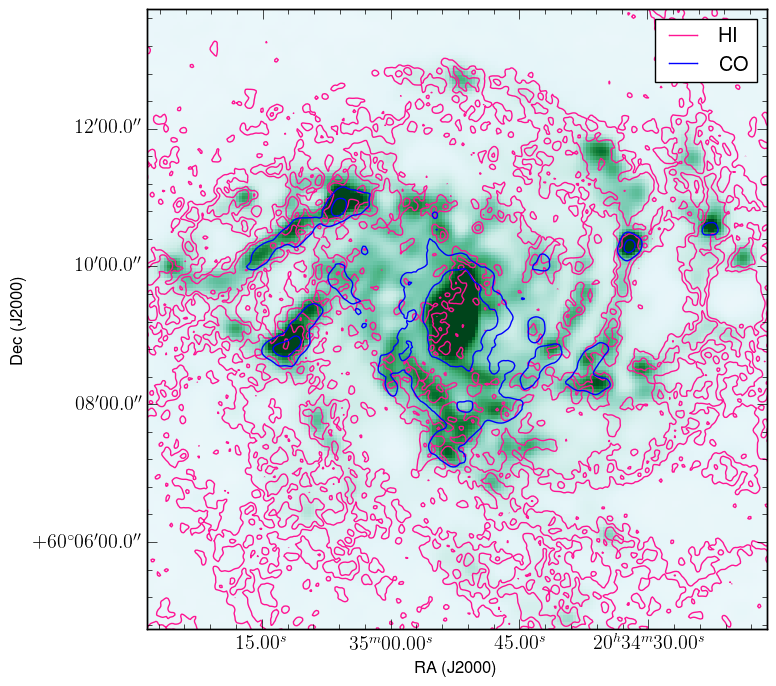}
\caption{Image of the galaxy NGC6946 at 160~$\mu$m with CO isophotes overlaid (blue contours). Red contours: left panel, 8~$\mu$m isophotes; right panel, HI isophotes. \hfill} 
\label{NGC6946_CO_HI}
\end{figure}


\section{OBSERVATIONS AND DATA REDUCTION}

We consider here a sample of galaxies that were
each included in a number of observational surveys: THINGS\footnote{http://www.mpia.de/THINGS/Data.html}\cite{THINGS} (HI line at 21~cm, Karl G. Jansky Very Large Array, VLA), KINGFISH\footnote{http://herschel.esac.esa.int/Science\_Archive.shtml}\cite{kingfish} (FIR region, 70, 100, and 160~$\mu$m, Herschel Space Observatory), SINGS\footnote{http://sings.stsci.edu}~\cite{sings} (NIR and MIR regions, 3.6, 4.5, 5.8, 8.0 and 24~$\mu$m,  Spitzer Space Telescope) and HERACLES~\cite{heracles}\footnote{http://www.cv.nrao.edu/$\sim$aleroy/heracles\_data/} (CO (2--1) line, IRAM). We selected 11 galaxies: Holmberg II, IC 2574, DDO 154, NGC 628, NGC 925, NGC 2976, NGC 3351, NGC 3627, NGC 4736, NGC 5055, NGC 6946. These belong to various morphological types, and so have a variety of metallicities, star-formation rates, etc. Table~1 presents information on these galaxies, namely, the equatorial coordinates of their centers \cite{deVaucoul91}, distances (with references), morphological types \cite{deVaucoul91}, metallicities averaged over the galaxy \cite{moustakas2010} (PT05 calibration), and the numbers of selected regions. The numbers in parentheses show the number of regions where we could determine the dust parameters by means of fitting their spectra (see below). The diameters of the regions are at least $12^{\prime\prime}$, which corresponds to the worst resolution of the images used (Herschel Space Observatory at 160~$\mu$m). We selected regions that could be visually distinguished in at least one of the following spectral ranges: 8~$\mu$m, 24~$\mu$m, 160~$\mu$m, 21 cm, and CO (2--1). Most of the regions are visible in some way in all of the images, but some emit in only one spectral range. To reduce selection effects, we included such regions in our sample. In total, about 300 regions were selected.

\begin{figure}
   \centering
  \includegraphics[width=13cm, angle=0]{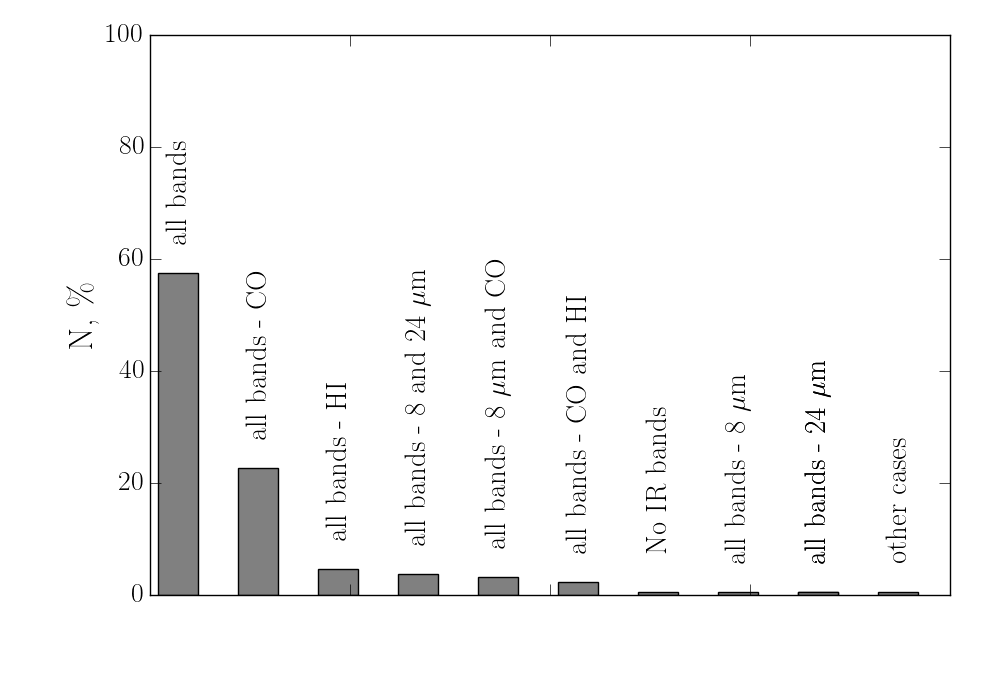}
   \caption{The percentage of objects (relative to the full sample) that match various conditions, e.g., the objects that demonstrate emission (${>}2\sigma$) in all the used ranges, objects that do not show only CO emission etc. \hfill} 
   \label{hist}
   \end{figure}

Figure ~\ref{NGC6946_CO_HI} displays an image of the galaxy NGC6946 at 160~$\mu$m with 8~$\mu$m and CO isophotes overlaid. Figure 1b shows the same image at 160~$\mu$m with HI and CO isophotes. These images show that regions that are bright at 160~$\mu$m virtually always coincide with regions emitting at 8~$\mu$m, while CO emission coincides only with the largest and brightest SFRs. If CO isophotes corresponding to lower brightnesses are included, this demonstrates that CO molecules are present virtually everywhere where dust is present, but the CO regions do not display such sharp contours as the dust; i.e., the distribution of CO molecules in the galaxy is more diffuse than that of dust. Unlike dust and molecular gas, atomic gas is virtually ubiquitous. This gas is more clearly visible in spiral arms, but is not necessarily associated with SFRs; in addition, it is abundant at the peripheries of galaxies. This picture is expected: the abundance of molecular gas should be enhanced mainly in SFRs, while the range of conditions for the existence of atomic hydrogen is much wider, and atomic hydrogen can exist both inside SFRs (around HII regions) and separate from them. Nevertheless, in spite of the spatial separation of the atomic hydrogen, molecular hydrogen, and dust, the studied regions are large enough to enable us to take into account all three components and compare their properties.

Figure \ref{hist} shows the fractions of objects in our sample based on their emission in specific bands. We took an object to emit in a band if the signal-to-noise ratio in the band exceeded two. More than half of the selected objects show emission in all the bands. Slightly more than $20\%$ of the objects do not emit in the CO line (at the given CO-flux limit). It is possible that the molecular gas in these SFRs is depleted or the amount of remaining gas is insufficient to produce bright CO emission. About $4\%$ of the objects have no HI emission. Such regions probably contain molecular and ionized hydrogen, while the amount of atomic hydrogen is very small. These are most likely dense molecular clouds; however, judging from the presence of hot dust, stars have already formed in these clouds. Approximately $3\%$ of the objects do not display any emission at 8 and 24~$\mu$m, i.e., no emission from hot, small dust grains and PAHs. These objects may be either old SFRs, where small dust grains have been mostly destroyed, or young SFRs, where there are not enough stars heating the dust to produce this emission. In approximately $2\%$ of the objects, there is no emission at 8~$\mu$m and in the CO line. The PAHs and CO molecules in these objects have presumably been destroyed by UV emission.

Our sample includes some single objects that, for example, do not emit in both HI and CO lines, or for which there is no dust emission, but since we did not aim to classify the objects in our sample, we do not consider these objects individually. Such objects in some way deviate from the ``norm'' (which we take to be the presence of emission in all bands), and could therefore affect our results, deviating from the obtained relationships.

The images at the different wavelengths were taken with different angular resolutions. To reduce
all the IR images to a common angular resolution, equal to the 160-$\mu$m resolution, we convolved all the images with the convolution kernel taken from \cite{2011PASP..123.1218A}. In addition, all the IR and COimages were rescaled to a common pixel size, equal to $2.85''$ (the pixel size for the 160-$\mu$m images). 
Finally, all the IR images were converted to a common intensity unit, Jy/pixel. After this preliminary preparation of the observational data, we performed aperture photometry of the selected regions, subtracting the background. The procedures for the photometry and estimation of the measurement errors are described in \cite{pahcycle}. The background was always estimated using a six-pixel-wide ring surrounding the studied region. Given the pixel size for the IR images, the width of the background ring is approximately twice the aperture size. The use of such a wide ring made it possible to smooth adjacent bright regions and correctly take into account the diffuse background in the aperture.

Further, we will pay considerable attention to the PAH and VSG emission at 8 and 24~$\mu$m, 
respectively. However, a significant fraction of the emission at these wavelengths is generated by stars. In addition, emission at 8~$\mu$m  is produced not only by PAHs, but also by larger dust grains. In order to distinguish only the PAH emission at 8~$\mu$m ($F_{8}^{\textrm{afe}}$) and only the VSG emission at 24~$\mu$m ($F_{24}^{\textrm{ns}}$),we applied the method for estimating the stellar and large-grain contributions at 8~$\mu$m described in~\cite{marble2010}.

The results of our aperture photometry for one of the galaxies, Holmberg II, are presented in Table 2. Analogous data for the rest of our galaxies are available in electronic form. Table 2 gives the SFR positions, aperture sizes, and flux densities at all the wavelengths considered. For the 8~$\mu$m and 24~$\mu$m bands we present both the total measured fluxes $F_8$ and $F_{24}$ and the corrected fluxes $F_{8}^{\textrm{afe}}$ and $F_{24}^{\textrm{ns}}$. Further, for conciseness,we will use  $F_8$ and $F_{24}$  to refer to the corrected fluxes.

\begin{figure}
\includegraphics[width=7cm, angle=0]{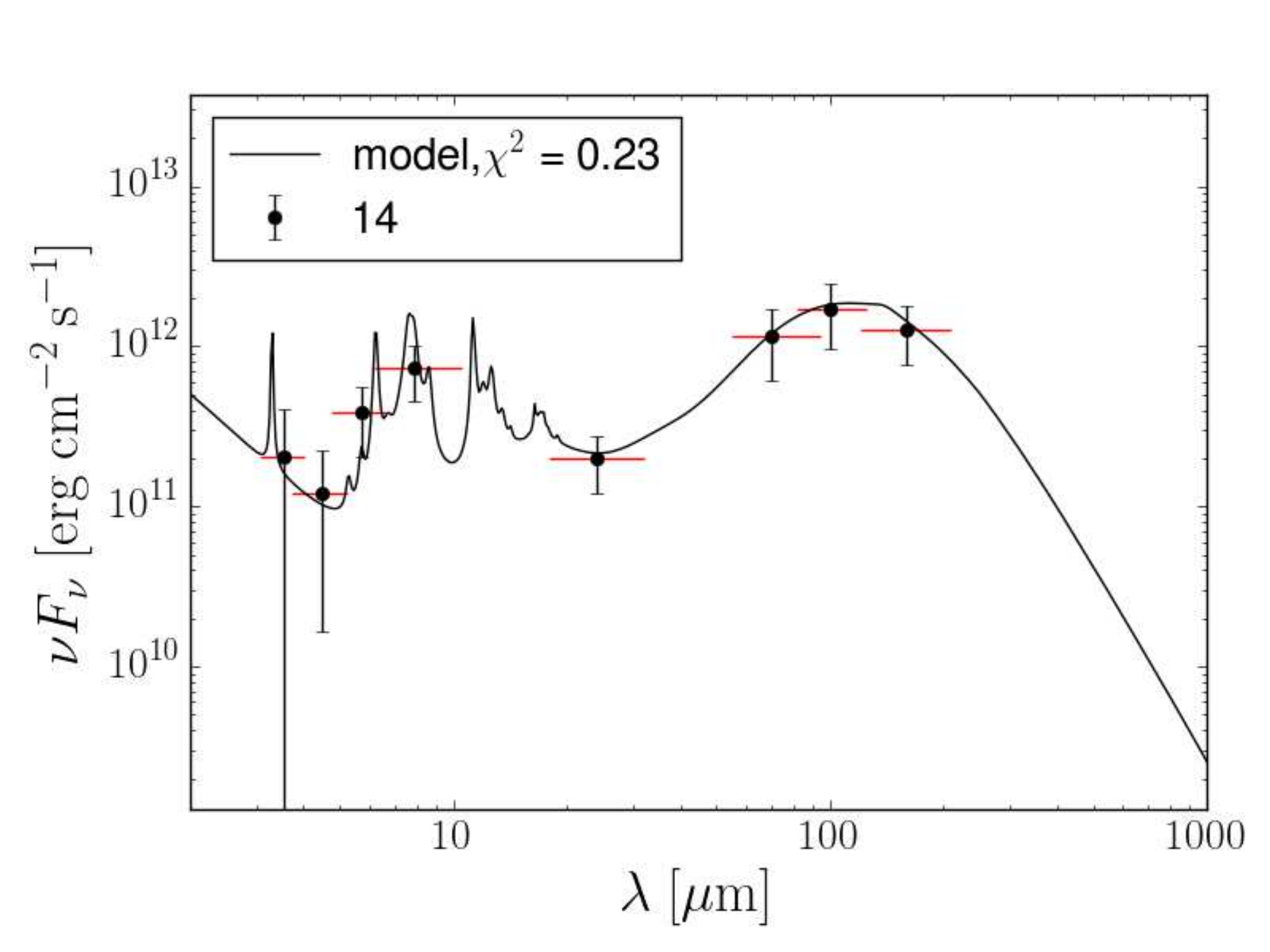}
\includegraphics[width=7cm, angle=0]{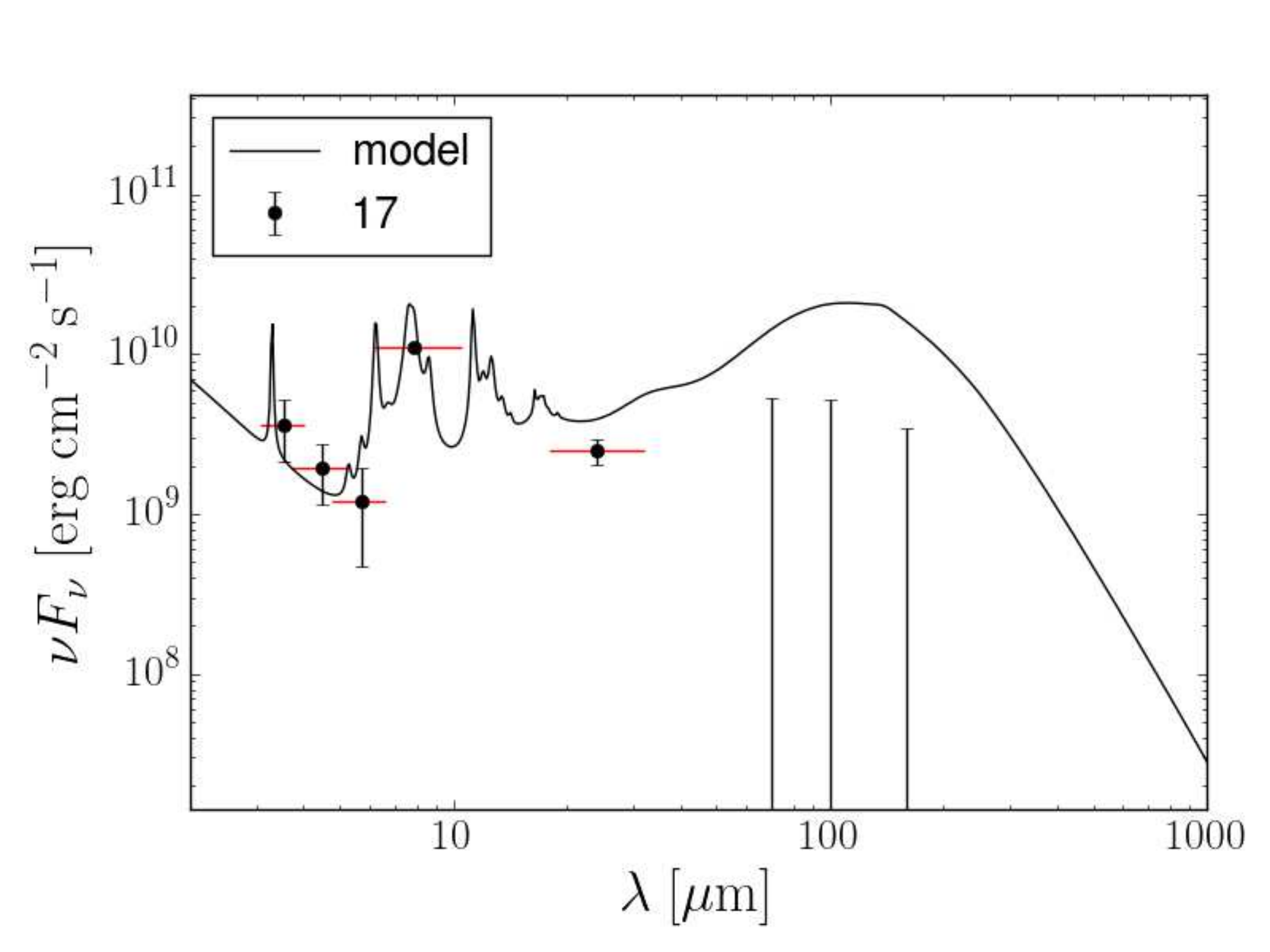}
\caption{Example spectra with good and bad fits. Left panel: region 14 in NGC~5055; right panel: region 17 in IC~2574, respectively. \hfill} 
\label{fitting}
\end{figure}

The continuum photometric data were used to determine the parameters of the dust and radiation fields in the model of Draine and Li \cite{2007ApJ...657..810D}. This model assumes that a dust cloud with mass $M_{\textrm{dust}}$ and a fraction of PAHs $q_{\textrm{PAH}}$ is heated by a radiation field whose intensity $U$ lies between $U_{\textrm{min}}$ The dust mass fraction $(1-\gamma)$ is illuminated only by radiation with intensity $U_{\textrm{min}}$, and the mass fraction $\gamma$, by radiation whose intensity lies within the indicated limits and obeys power-law distribution with index $\alpha$, 
\begin{equation}
\frac{dM_{\rm dust}}{dU} =
(1-\gamma) M_{\rm dust} \delta (U-U_{\min})+
\gamma M_{\rm dust}\frac{\alpha-1}{U_{\min}^{1-\alpha}-U_{\max}^{1-\alpha}} U^{-\alpha}. 
\label{main_dife}
\end{equation}

Following~\cite{Draineetal2007}, we set $U_{\max}$ = $10^6$  and $\alpha$ = 2. Equation~(1) can be used to calculate the model spectrum of dust emission in the studied object. Fitting the model spectra to the aperture photometry data, we estimated the PAH mass fraction $q_{\textrm{PAH}}$, the total mass of dust $M_{\textrm{dust}}$, the minimum radiation intensity $U_{\textrm{min}}$, the volume  fraction of the region where radiation is enhanced $\gamma$, and the parameter $\Omega$ characterizing the NIR stellar contribution for each SFR. All the intensities are given in the units of the mean radiation field in the solar vicinity~\cite{mmp83}.

In the fitting, we used our spectral database~\cite{2007ApJ...657..810D}, where the PAH mass fraction $q_{\textrm{PAH}}$ is limited to the values $0.5{-}4.8\%$, and $U_{\textrm{min}}$, to values 0.1--25. The fitting applied $\chi^2$ minimization by Levenberg--Marquardt gradient method, where $\chi^2$ is the sum of squared differences between the observed $F_{\nu,{\textrm{band}}}^{\textrm{obs}}$ and model $F_{\nu,{\textrm{band}}}^{\textrm{model}}$ fluxes in each of the IR bands, taking into account the observational uncertainties ($\Delta F_{\textrm{band}}^{\textrm{obs}}$):

\begin{equation}
\chi^2=
\sum_{\textrm{band}}{\frac{(F_{\nu,{\textrm{band}}}^{\textrm{obs}}-F_{\nu,{\textrm{band}}}^{\textrm{model}})^2}{(\Delta
F_{\textrm{band}}^{\textrm{obs}})^2}}.
\end{equation}

Figure \ref{fitting} presents examples of spectral fits for two SFRs, one fitted well and the other fitted poorly. The solid horizontal lines show the widths of the IRAC, MIPS, and PACS photometric bands. Poorly fitted regions were excluded from further consideration. Usually, regions of poor fits had fluxes in some bands that were unphysically low. These cases are considered in the next section.

\subsection{Excluding of regions}

As was previously noted, we initially selected regions distinguishable in at least one of the spectral ranges considered. However, in order to carry out spectral fitting, we must have regions with meaningful fluxes in all the IR bands. Therefore, we had to exclude some of these. The numbers in parantheses in the last column of Table 1 denote the numbers of SFRs without taking into account excluded and poorly fitted regions.

{\textbf{NGC 6946.}} We excluded eight regions in this galaxy, five of which do not display observable fluxes in all the IR bands. In the remaining three regions, the fitting results were unphysical as a result of the presence of radiation from neighboring regions in the aperture.

{\textbf{NGC 5055.}} Five regions were excluded. Two of these were selected based on their CO emission, but do not emit in the IR. The fitting results for the remaining three were unphysical due to the presence of radiation from neighboring regions in the aperture.

{\textbf{NGC 4736.}} Four regions were excluded. In two of these, no radiation was registered at some of the wavelengths considered here. Two others have complex structures (which are also different in different IR bands), leading to the presence of radiation from neighboring regions in the aperture.

{\textbf{NGC 3627.}} Two regions were excluded because of the absence 
of emission at 24~$\mu$m.

{\textbf{NGC 3351.}} Ten regions were excluded, six due to the absence of radiation in one or more IR bands, with three of these demonstrating only CO emission.

{\textbf{NGC 2976.}} Two regions were excluded. These have indistinct structures, leading to the presence of radiation from neighboring regions in the aperture.

{\textbf{NGC 925.}} Thirteen regions were excluded. In seven of
these, no radiation was observed in some IR bands.

{\textbf{NGC 628.}} Nine regions were excluded, four due to indistinct structures.

{\textbf{Ho II.}} Five regions were excluded due to the absence of radiation in one or more bands.

{\textbf{IC 2574.}} Five regions with indistinct structures were excluded.

\begin{figure}[t!]
\includegraphics[width = 8cm, angle=0]{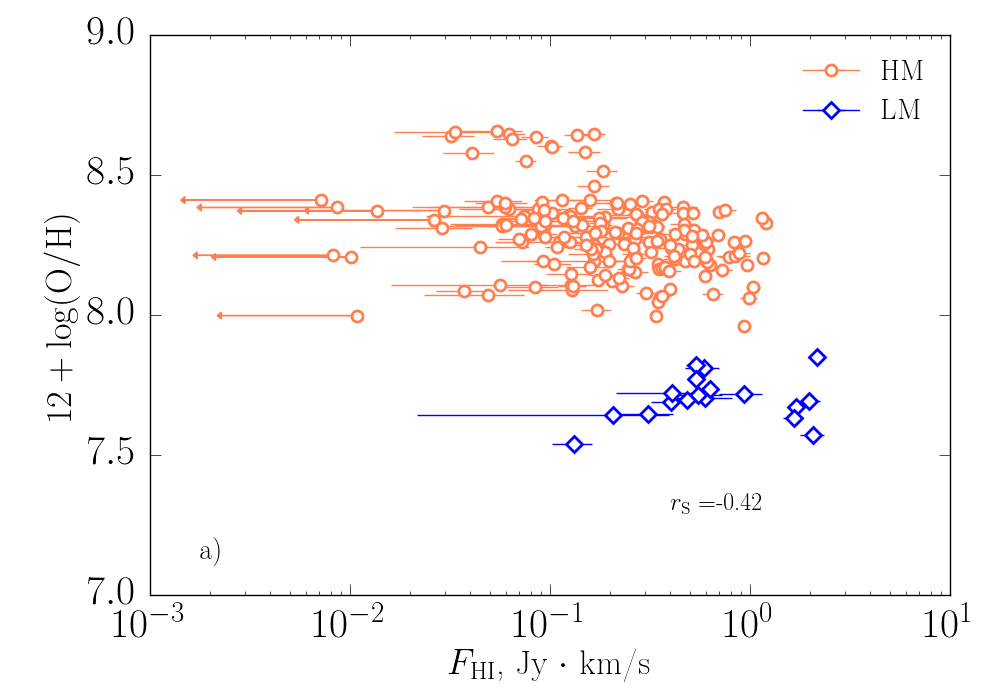}
\includegraphics[width = 8cm, angle=0]{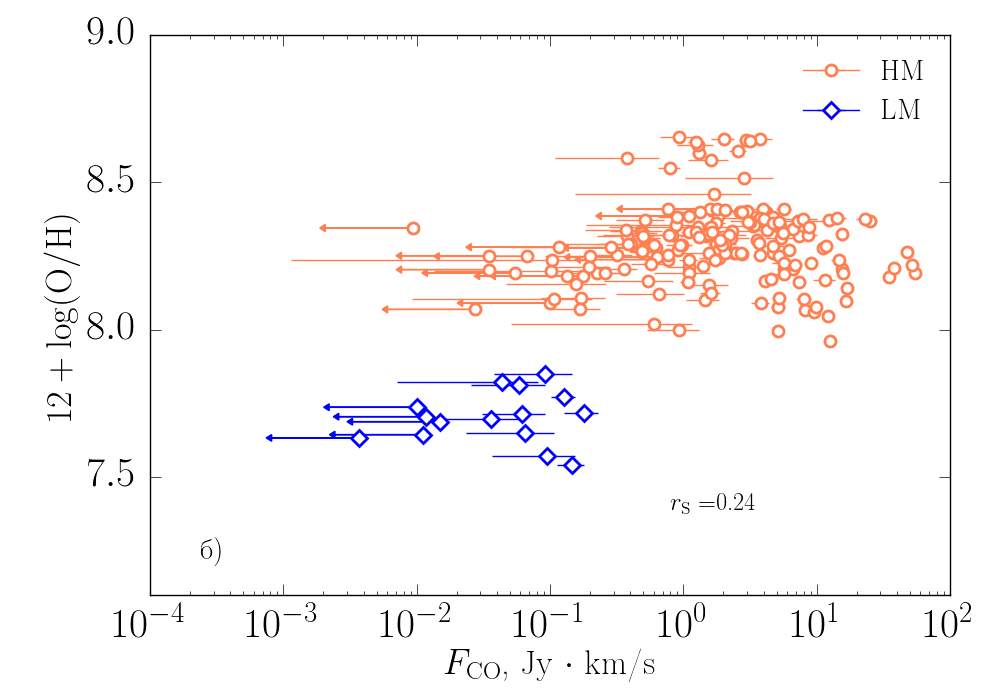}
\caption{Relationships between the fluxes in the HI (left) or CO (right) lines and SFR metallicities. Arrows here and below show the errorbars extending beyond the picture edges. \hfill.}
\label{COHI_Z}
\end{figure}

\subsection{Metallicities}

We calculated the metallicities of the selected regions in spiral galaxies approximately using the radial metallicity gradients obtained in \cite{moustakas2010}. We used metallicities estimated applying the method of \cite{PT05}. Metallicities in irregular galaxies were taken to be equal to their galaxy-averaged values. Since the metallicity scatter in irregular galaxies is small, this only slightly decreases the accuracy of our results; in addition, we focus on the differences between the properties of high-metallicity($12 +\log(\textrm{O/H})\geq7.9$) and low metallicity ($12 +\log(\textrm{O/H})<7.9$) regions,  while the precise metallicity values are not important for us. The metallicity only serves as a criterion for assigning a region to one of two broad groups.

\section{RESULTS}

We assumed that the abundances of different components of the ISMin SFRs are not independent,
so that their parameters should be correlated in some way. We considered both observed parameters ($F_8$, $F_{24}$, $F_{70}$, $F_{160}$, $F_{\textrm{CO}}$, and $F_{\textrm{HI}}$), and parameters derived from the fits to the spectral energy distributions ($M_{\textrm{dust}}$, $q_{\textrm{PAH}}$, $U_{\textrm{min}}$, and $\gamma$). In all the figures described below, we indicate the Spearman rank correlation coefficient $r_{\textrm{S}}$. Note that the derived parameters themselves are essentially functions of the measured IR fluxes.

We separately present and discuss the results for high-metallicity (HM) and low-metallicity (LM)
SFRs. Figure \ref{COHI_Z} shows the relationships between the metallicities of the studied regions and their fluxes in the atomic hydrogen and CO lines. The arrows in this and the following figures show error bars that extend beyond the edges of the figure panels. The galaxies form two isolated groups according to their metallicities (denoted LM and HM). There is no correlation between $12 + \log(\textrm{O/H})$ and the fluxes, but SFRs in LM galaxies are shifted to the right relative to SFRs in HM galaxies in Fig. \ref{COHI_Z}a. In other words, on average, the selected regions in LM galaxies are brighter in the 21-cm line. This could have two origins. First, if the masses of atomic hydrogen in all the selected SFRs are approximately the same, then the regions in LM galaxies may really be {\emph{brighter}}, e.g., due to higher temperatures. Second, we could have selected, on average, more {\textit{massive}} regions in the LM galaxies. When the metallicity is low, the amounts of dust and CO in low-mass regions may be so small that they are not detectable in the maps used.

In the latter case, we expect that the lower limit on the CO flux will be the same for both types of galaxies. In fact, Fig. \ref{COHI_Z}b shows that this is not the case: the minimum CO fluxes of selected regions in LM galaxies are appreciably lower than those in HM galaxies. If, on average, the selected regions in all the galaxies have comparable total masses, then the regions in LMgalaxies should indeed be less bright in the CO line.

Note that themetallicities in the sample of galaxies do not overlap; i.e., we cannot compare relatively metal-poor SFRs from metal-rich galaxies with relatively metal-rich SFRs from metal-poor galaxies.

\begin{figure}[t!]
\includegraphics[width=0.45\textwidth]{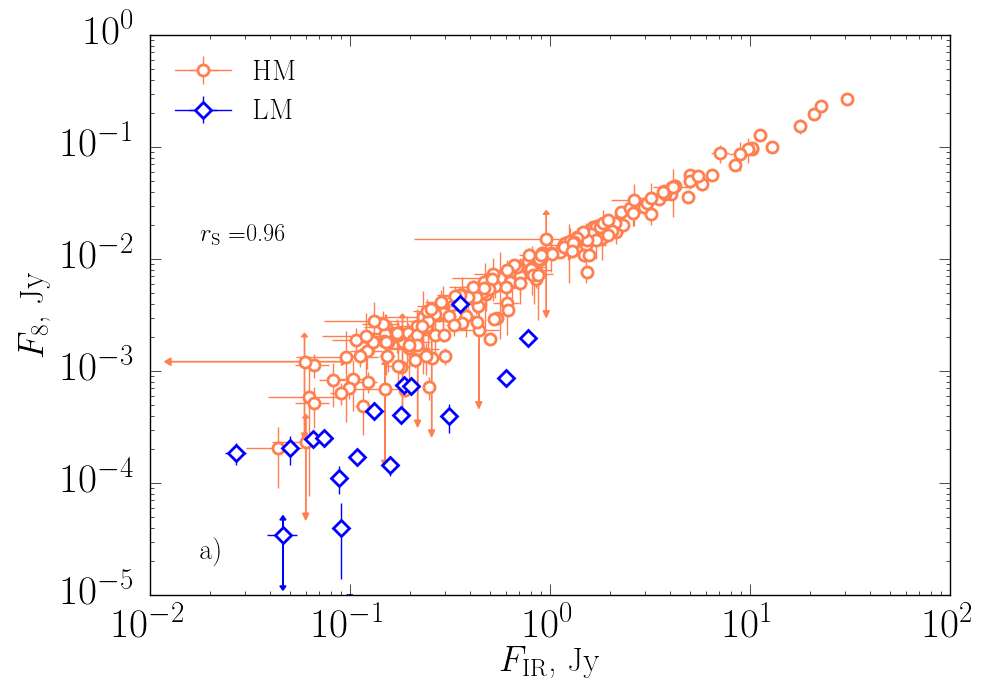}
\includegraphics[width=0.45\textwidth]{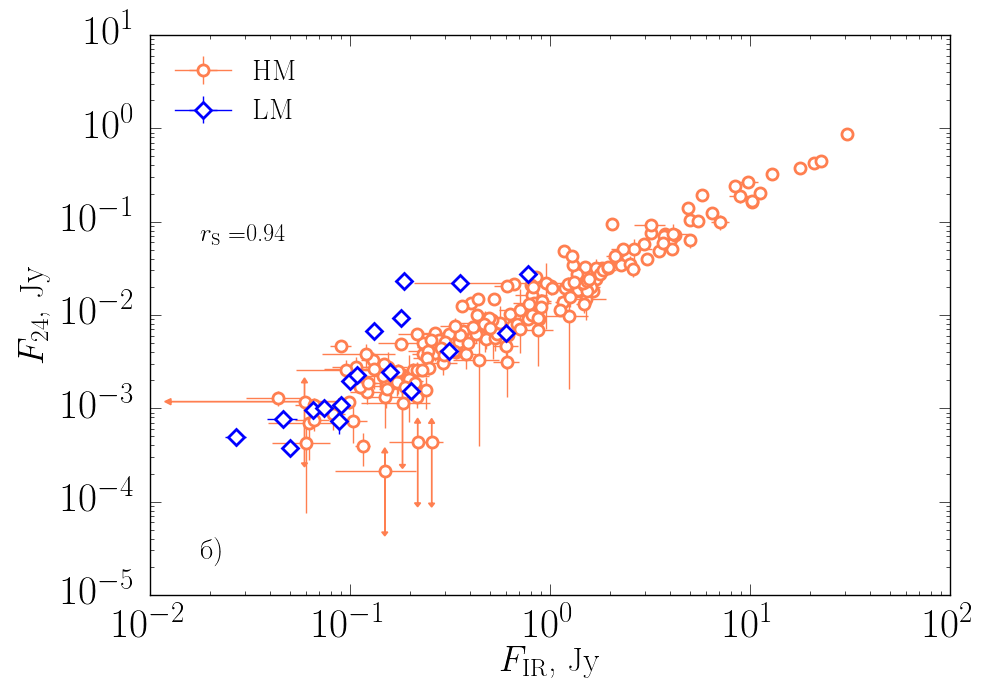}
\caption{Comparison of the full fluxes in the FIR range with the 8~$\mu$m (left) and 24~$\mu$m (right) fluxes. \hfill}
\label{obsfluxes1}
\end{figure}

\begin{figure}[t!]
\includegraphics[width=0.45\textwidth]{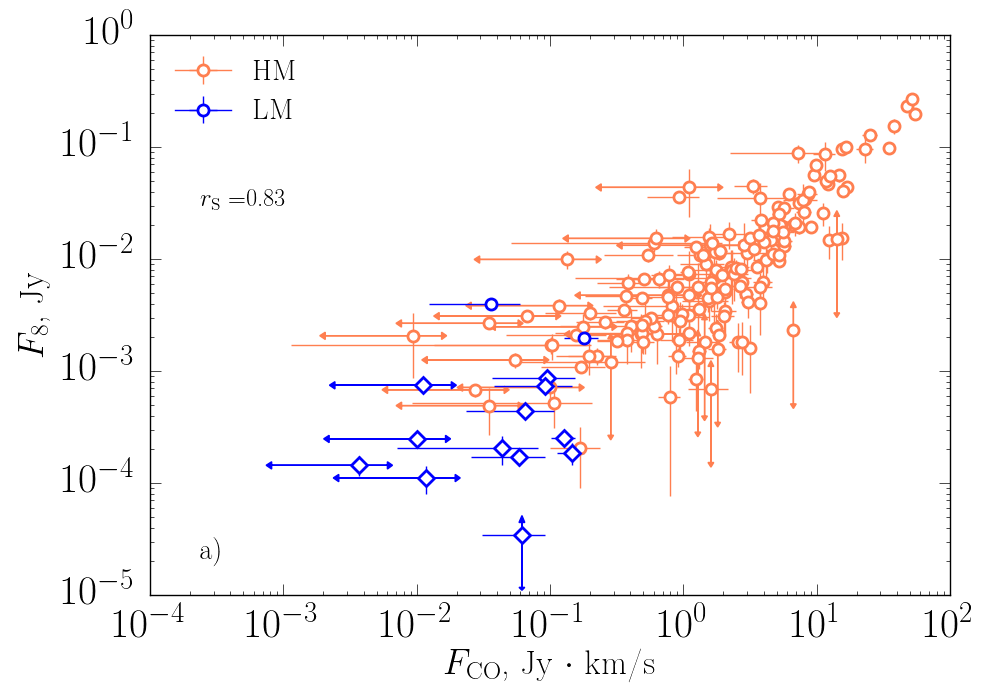}
\includegraphics[width=0.45\textwidth]{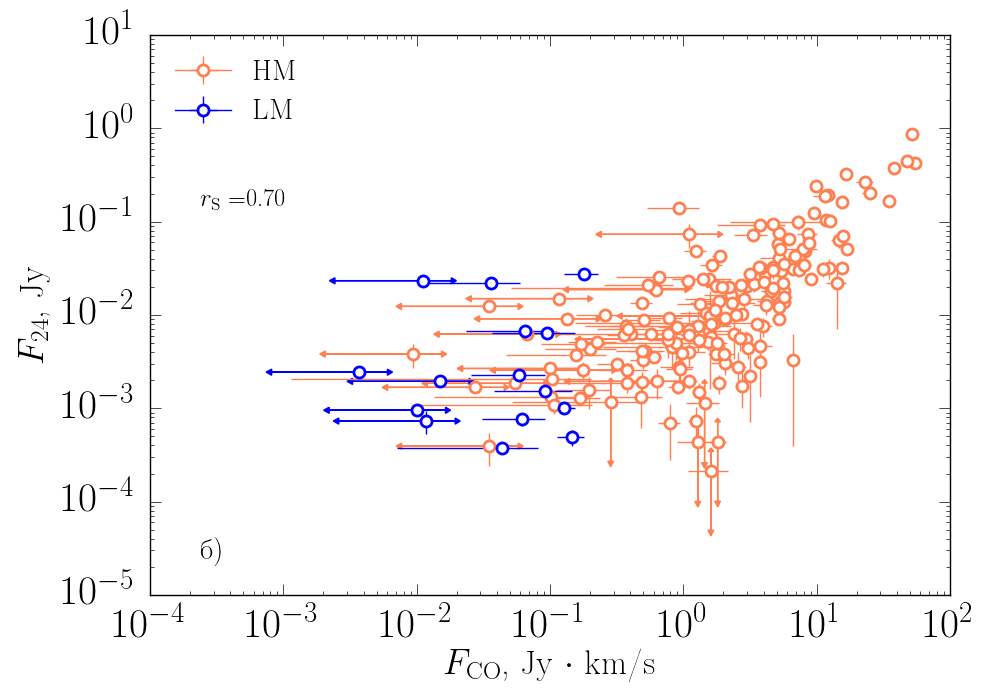}
\caption{Comparison of CO fluxes with the 8~$\mu$m (left) and 24~$\mu$m (right) fluxes.
\hfill}
\label{obsfluxes2}
\end{figure}

\subsection{Observed fluxes}

Let us first consider the correlations between fluxes in different spectral ranges. Figure ~\ref{obsfluxes1} shows how are the 8 and 24~$\mu$m  fluxes are related to the total FIR flux (here and below $F_{\textrm{IR}}=F_{70}+F_{100}+F_{160}$). Correlations are visible in both diagrams; however, in Fig. ~\ref{obsfluxes1}a, they are different for galaxies with different metallicities, namely, the same FIR luminosities correspond to lower 8~$\mu$m fluxes in LM than in HM galaxies. This relationship represents the known relative weakening of PAH emission in LM galaxies and SFRs. The 24~$\mu$m fluxes are correlated with $F_{\textrm{IR}}$ independent of the SFR metallicities, confirming the conclusion of \cite{khramtsova1} that the contribution of the 24~$\mu$m emission to the total IR luminosity does not depend on the metallicity.

Figure~\ref{obsfluxes2} shows the correlations between the 8 and 24~$\mu$m fluxes  and the CO (2--1) flux. Both correlations are significant; however,  even here, there are some differences between the 8~$\mu$m and 24~$\mu$m fluxes. Points in Fig.~\ref{obsfluxes2}a corresponding to SFRs with different metallicities lie approximately along the same line, indicating a common correlation between $F_8$ and $F_{\textrm{CO}}$ for HM and LM SFRs. In this respect, the CO emission and 8~$\mu$m are related. The CO and 24-$\mu$m fluxes are also correlated, but points corresponding to LM SFRs are shifted to the left in Fig.~\ref{obsfluxes2}b, i.e., a given 24~$\mu$m flux corresponds to a lower CO flux in the case of LM SFRs than of HM SFRs.

\begin{figure}[t!]
\includegraphics[width=0.45\textwidth]{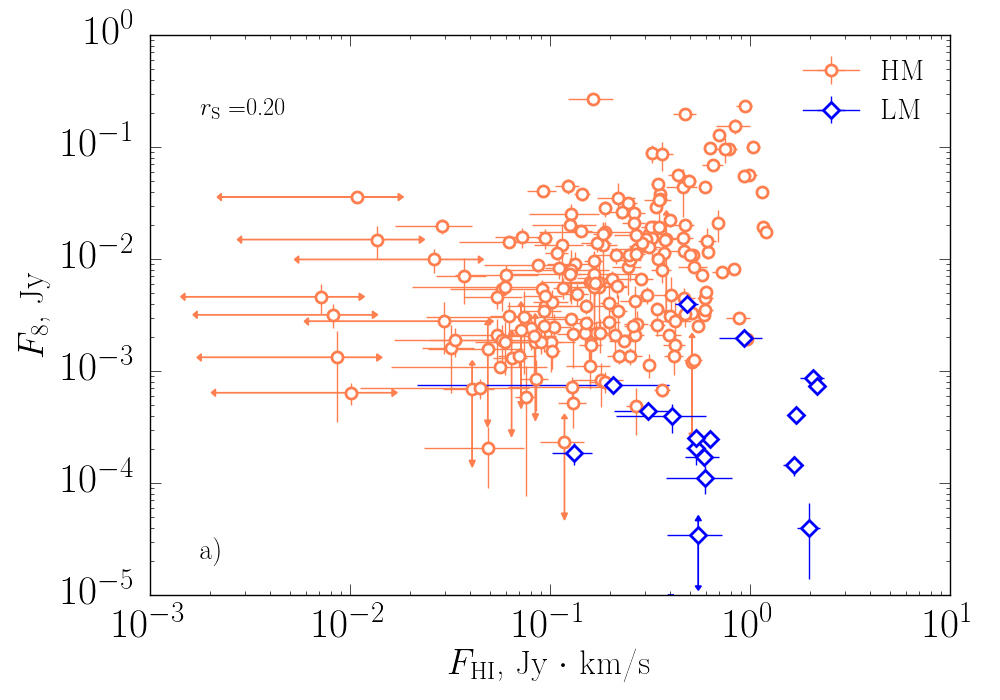}
\includegraphics[width=0.45\textwidth]{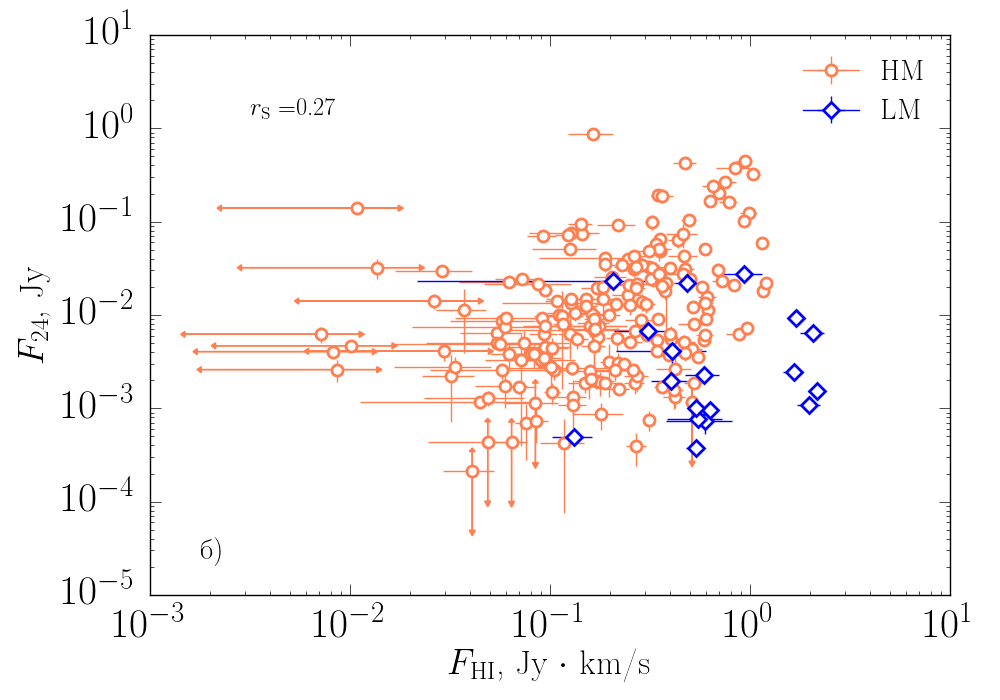}
\includegraphics[width=0.45\textwidth]{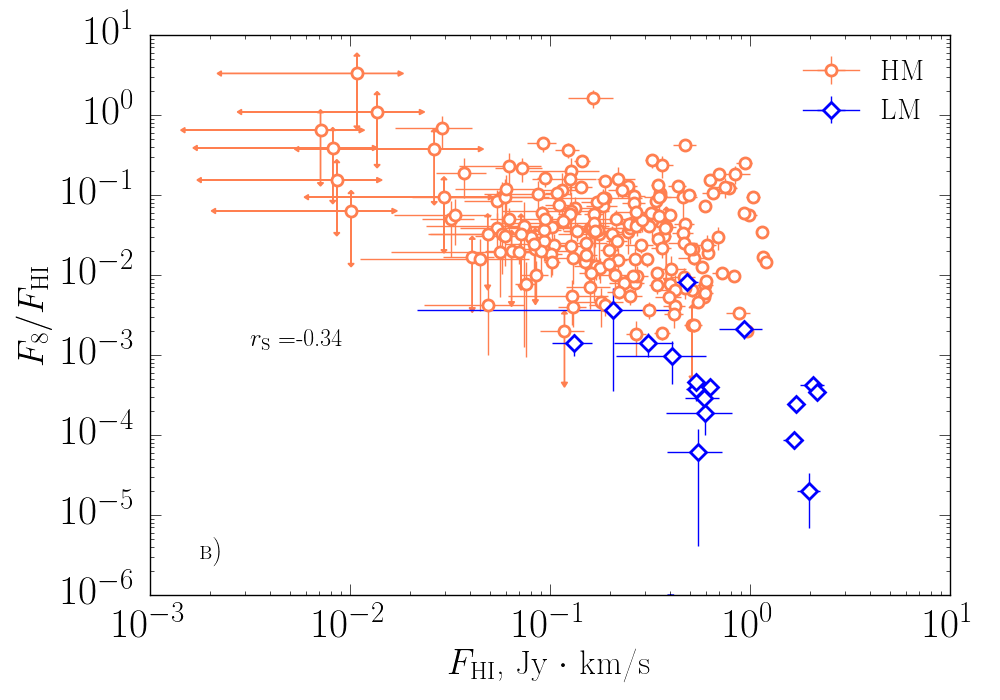}
\includegraphics[width=0.45\textwidth]{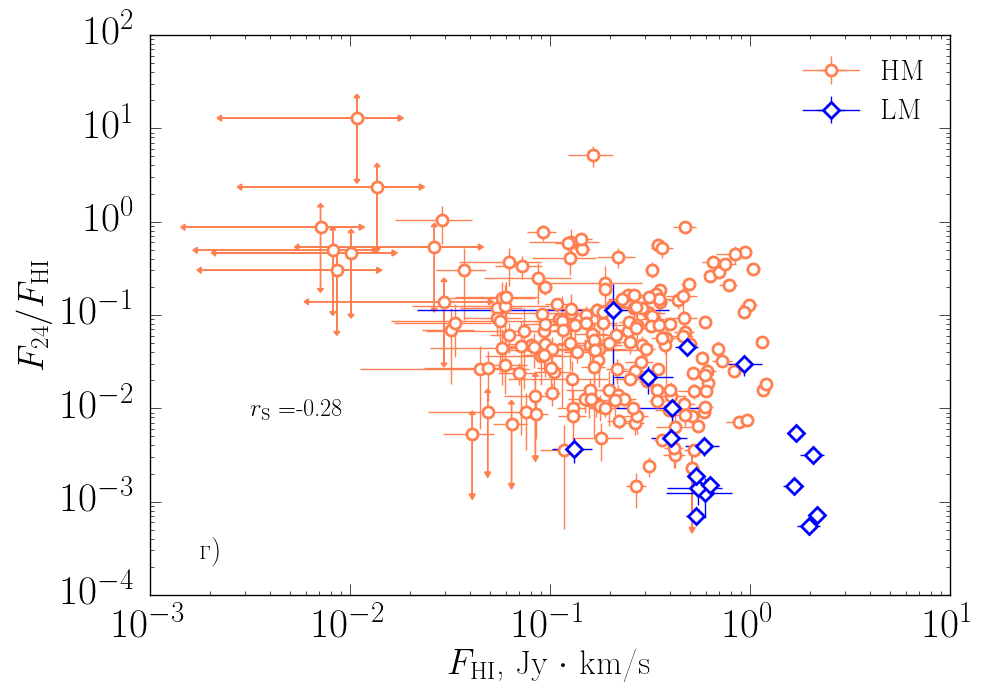}
\includegraphics[width=0.45\textwidth]{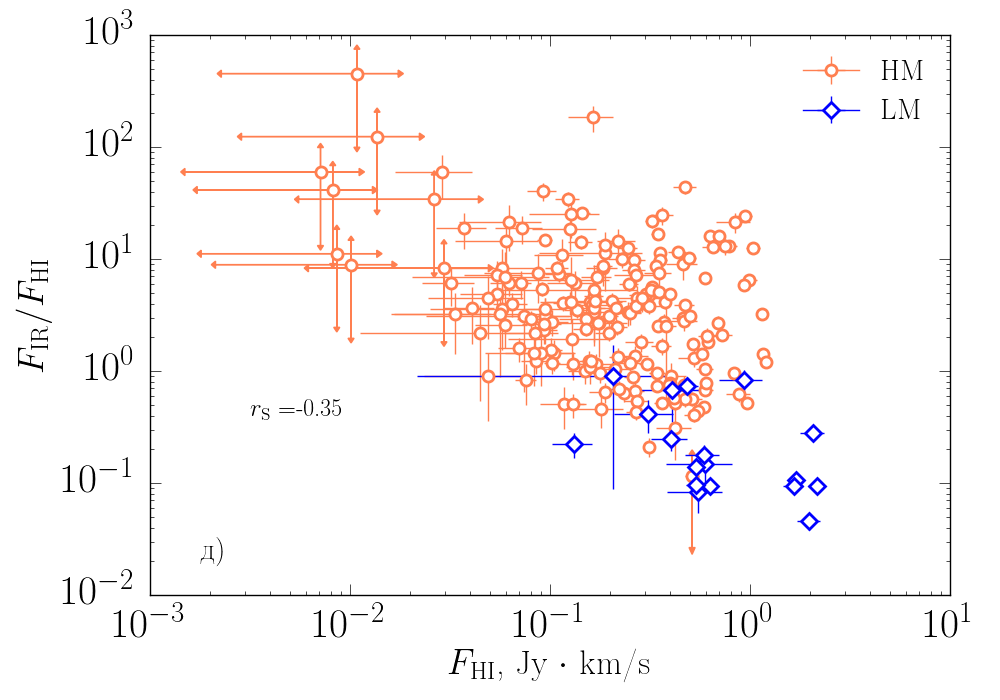}
\includegraphics[width=0.45\textwidth]{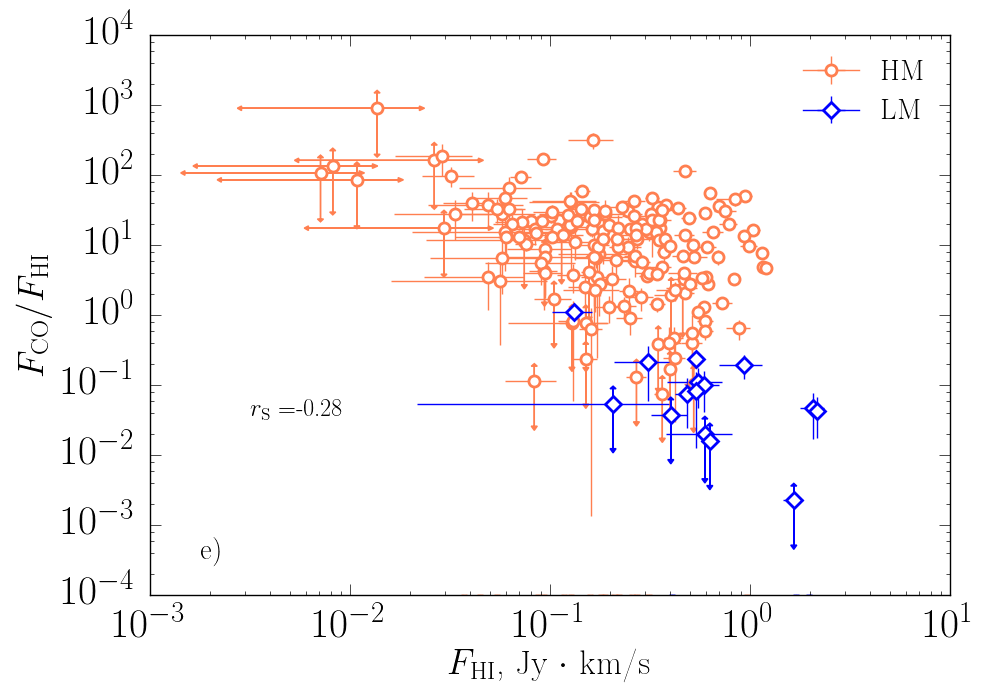}
\caption{Relationships between 8-$\mu$m (upper left panel), 24-$\mu$m (upper right panel) and HI fluxes. Plots in the middle raw demonstrate correlations between HI fluxes and the ratios $F_8/F_{\textrm{HI}}$ and $F_{24}/F_{\textrm{HI}}$; in the lower raw, between HI fluxes and the ratios $F_{\textrm{IR}}/F_{\textrm{HI}}$ and $F_{\textrm{CO}}/F_{\textrm{HI}}$. \hfill}
\label{obsfluxes3}
\end{figure}

The 21-cm flux does not correlate with any of the other fluxes considered. As an example, Figs. ~\ref{obsfluxes3}a, ~\ref{obsfluxes3}b compare the HI fluxes with the fluxes at 8 and 24~$\mu$m. Some correlation can be seen for HM SFRs, in the sense that regions luminous at 21 cm are brighter at both 8 and 24~$\mu$m. Furthermore, LM and HM SFRs are separated in the plot for 8~$\mu$m, i.e., there is a clear boundary between them. There is no such clear boundary in the plot for 24~$\mu$m, where LMSFRs overlap with HM SFRs, and probably demonstrate the same correlation as HM SFRs; i.e., the brightness at 24~$\mu$m increases with the brightness at 21 cm in both 24~$\mu$m increases with the brightness at 21 cm in both types of SFRs. This again emphasizes the weaker dependence of the 24~$\mu$m emission on the metallicity.

Figures ~\ref{obsfluxes3}c, ~\ref{obsfluxes3}d display the correlations between
the HI flux and the ratios $F_8/F_{\textrm{HI}}$ and 
$F_{24}/F_{\textrm{HI}}$, the picture looks clearer here. In both cases, we can see an anticorrelation between the normalized IR fluxes and the HI fluxes: the higher the 21 cm flux, the lower the relative contribution of the IR bands. The difference between the HMand LM SFRs may reflect the trend noted above: for some reason, the SFRs we have selected in LM galaxies are brighter in the HI line (although the brightness of the HI emission was not the main selection criterion). However, we can now see the correlation within groups of SFRs with different metallicities as well. The correlation shows that the 8 and 24~$\mu$m emission is relatively brighter in regions with lower 21 cm fluxes. This may be an effect of the SFR ages: when the conditions for generating MIR and NIR emission arise in SFRs, the amount of atomic hydrogen becomes low. The plots for the total IR and CO fluxes in Figs. ~\ref{obsfluxes3}e, ~\ref{obsfluxes3}f look similar, which can likewise can be explained in terms of age: the more molecular gas and dust in a SFR, the less HI it contains. This possible influence of age on the
observed fluxes hinders our interpretion of the data: essentially, the age is an additional parameter that cannot be estimated based on the available data.

\subsection{Derived parameters}

In this subsection, we consider the relationships between the observed parameters of the SFRs and their parameters estimated from fitting their IR spectra. The derived SFR parameters are the mass of dust $M_{\textrm{dust}}$, the minimum intensity of the radiation field, $U_{\textrm{min}}$, and the mass fraction of dust illuminated by radiation with intensity higher than $U_{\textrm{min}}$, $\gamma$. We suppose that $U_{\textrm{min}}$ characterizes the total number of stars in the SFRs, and $\gamma$ the mass of dust that is in the vicinities of stars, and is therefore illuminated by more intense and probably harder radiation. We also estimated the masses of atomic and molecular gas,($M_{\textrm{HI}}$ and $M_{{\textrm{H}}_2}$ using Eq. (3) from \cite{THINGS}, and estimated $M_{{\textrm{H}}_2}$ using Eqs. (3) and (4) from \cite{Tanetal2013}.

\begin{figure}[t!]
\includegraphics[width=0.45\textwidth]{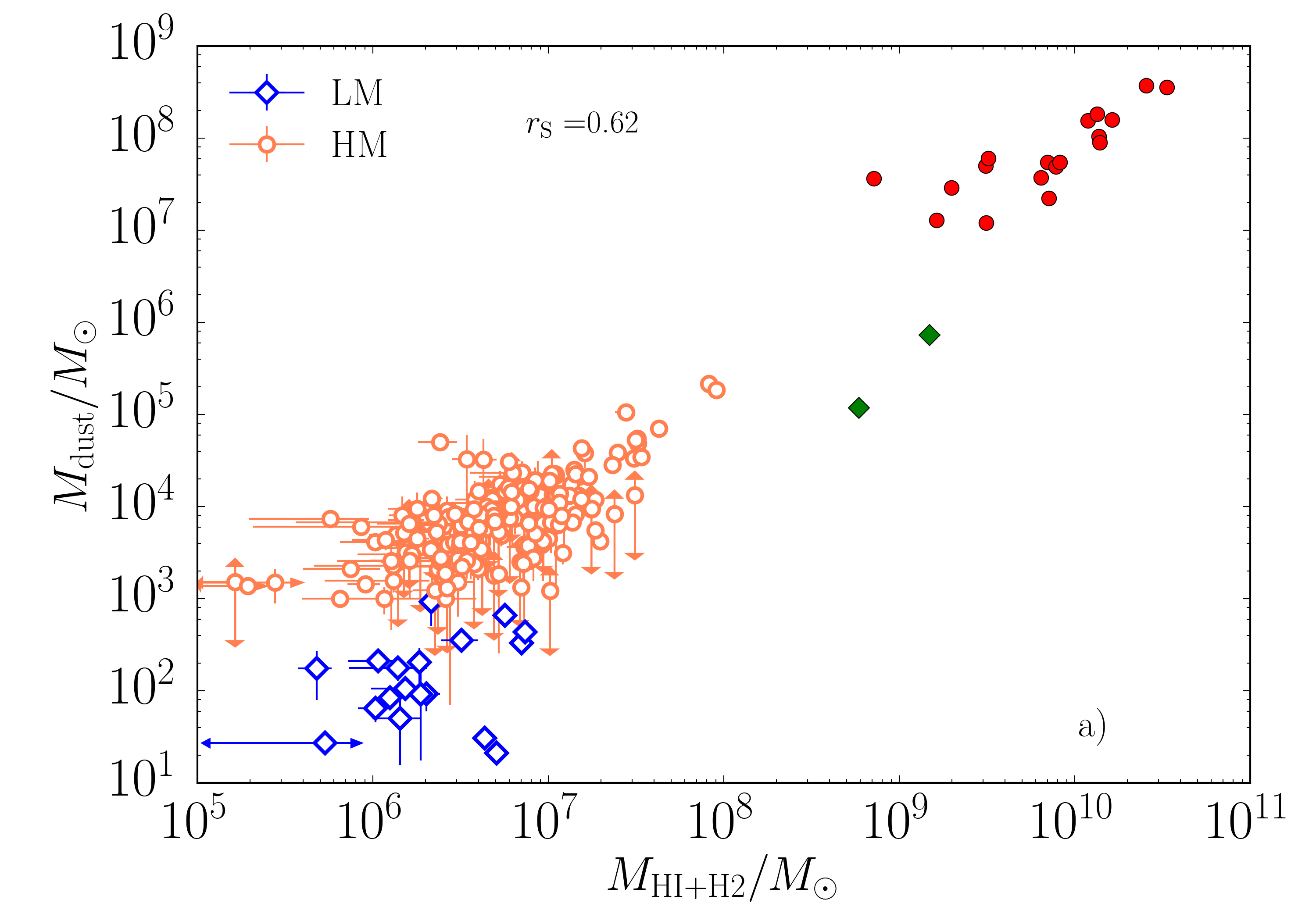}
\includegraphics[width=0.45\textwidth]{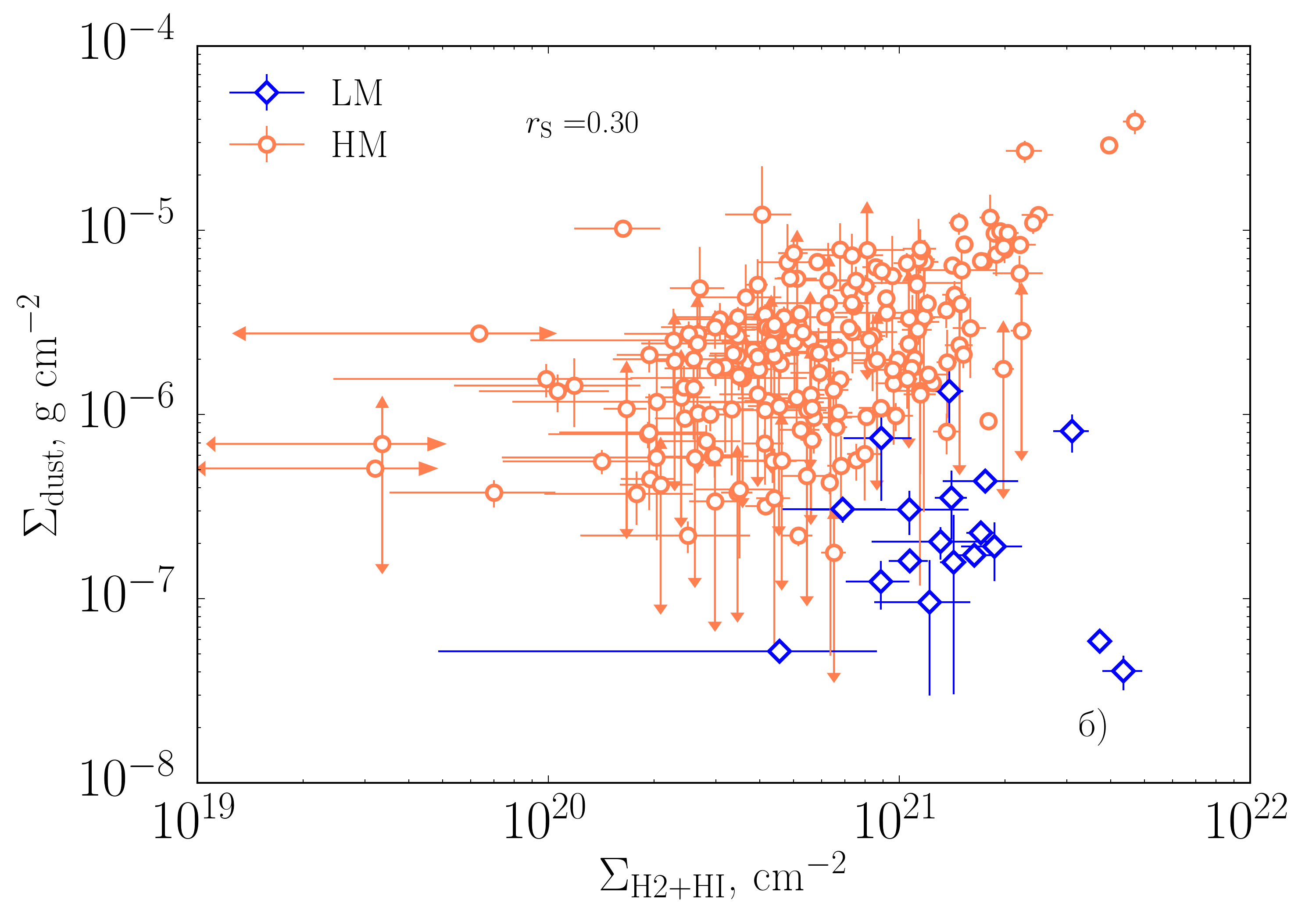}
\caption{Relationships between the masses (left) and surface densities (right) of gas and dust. Red and green symbols on the left diagram show the total masses of gas and dust in HM and LM galaxies, respectively, taken from~\cite{Draineetal2007}. \hfill}
\label{sanitycheck}
\end{figure}

\begin{figure}[t!]
\includegraphics[width=0.6\textwidth]{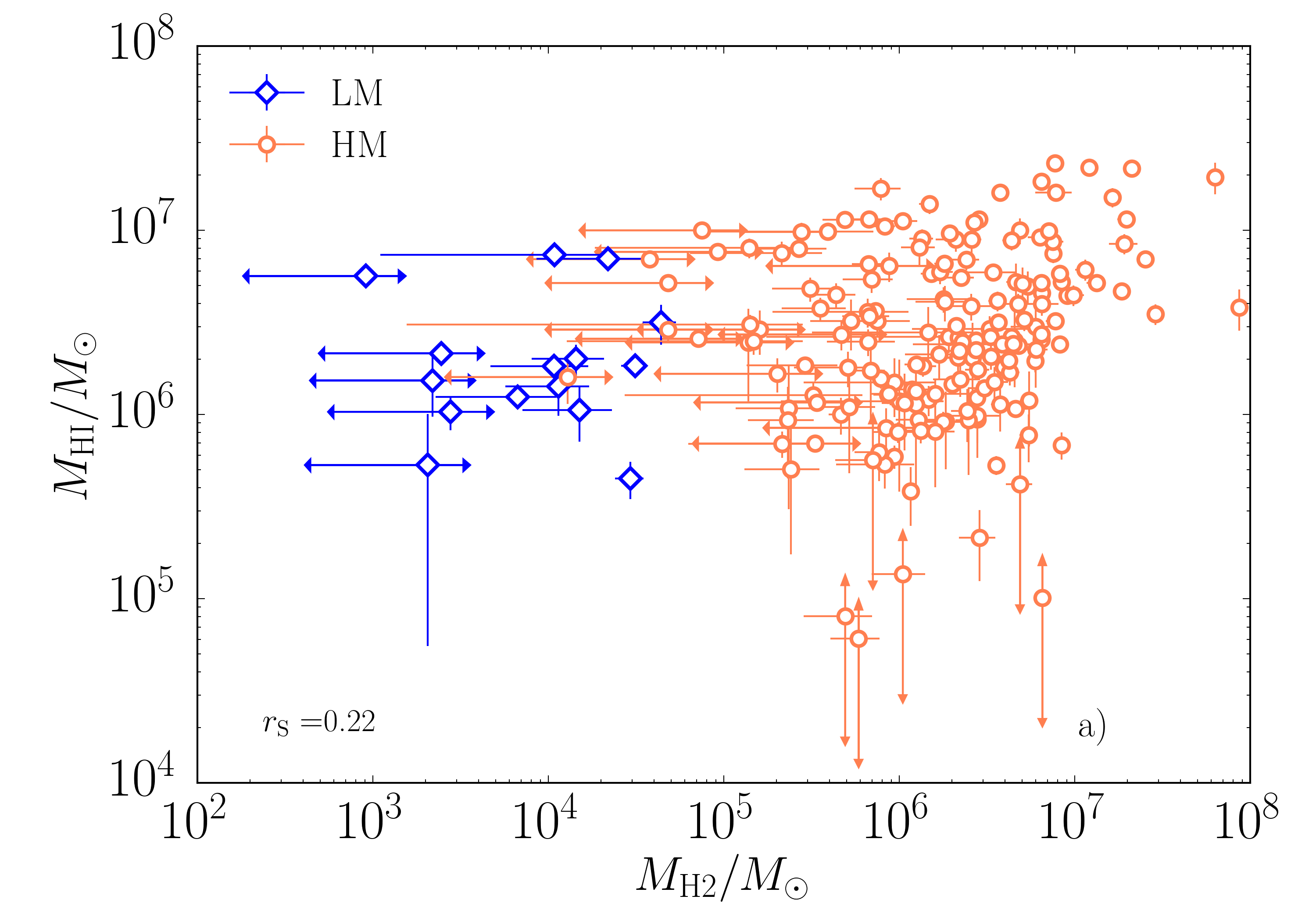}
\includegraphics[width=0.6\textwidth]{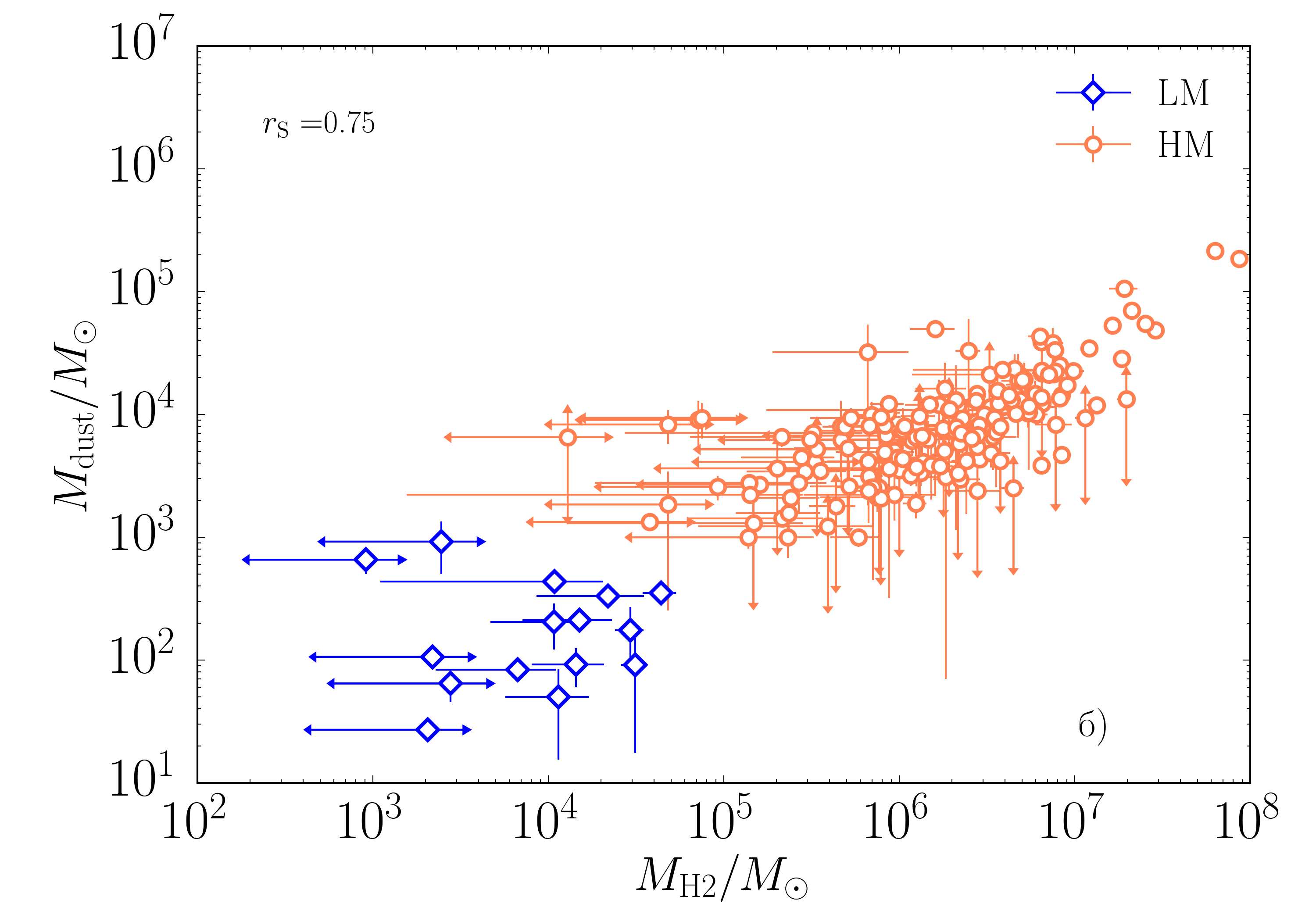}
\includegraphics[width=0.6\textwidth]{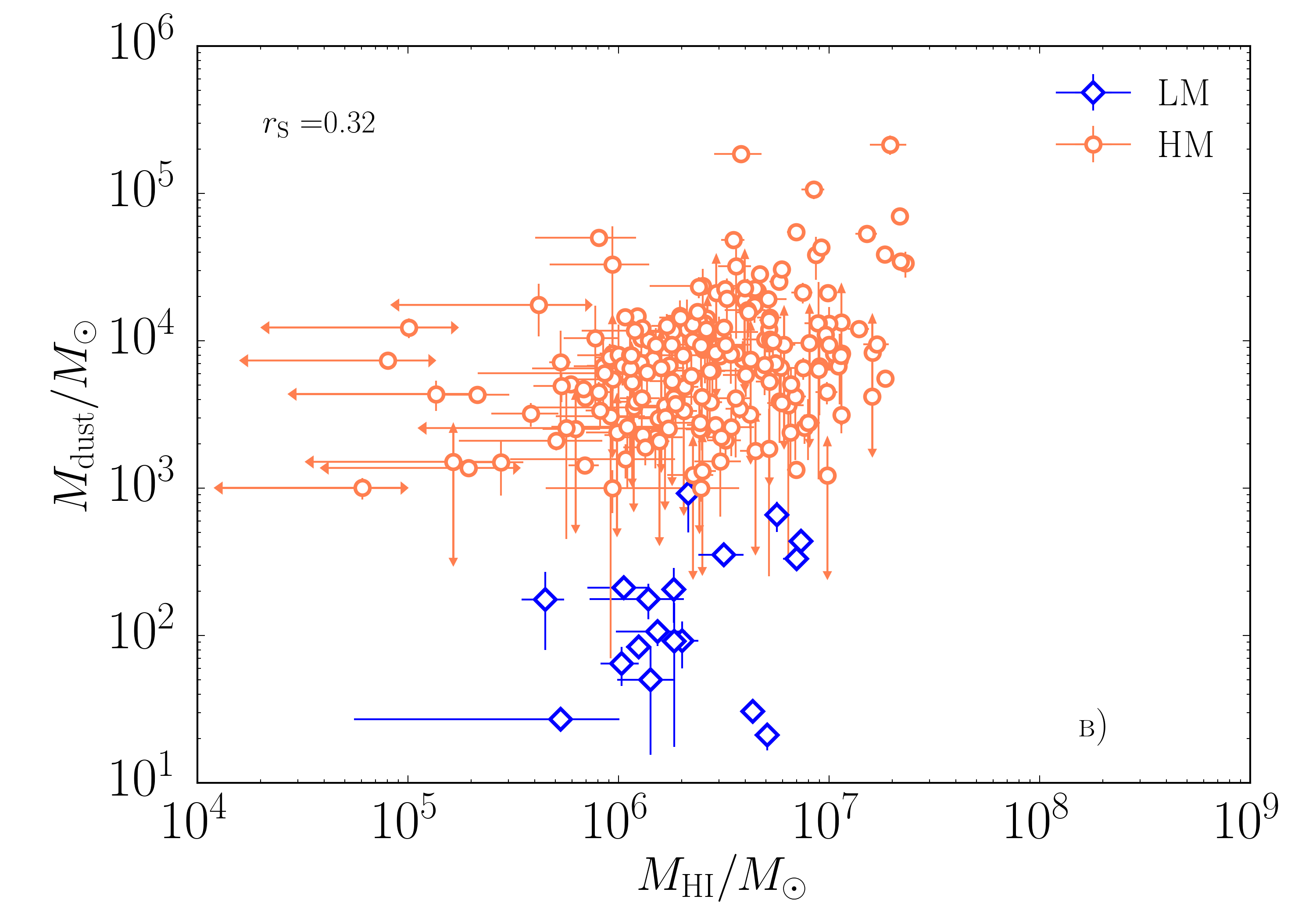}
\caption{Relationships between the masses of molecular and atomic hydrogen~(a), of molecular hydrogen and dust~(b), and of atomic hydrogen and dust~(c). \hfill}
\label{mgas_mdust}
\end{figure}

Our estimates of the masses and surface densities of gas and dust are compared in Fig.~\ref{sanitycheck}. For comparison, we also show in Fig. ~\ref{sanitycheck}a the total masses of gas and dust in galaxies from the SINGS survey \cite{Draineetal2007}. The HM and LM galaxies are denoted by red and green symbols, respectively. Our results are clearly consistent with those of \cite{Draineetal2007}: the dust masses are approximately proportional to the gas masses in SFRs of both metallicity groups, but the points corresponding to LM SFRs are, on average, lower then the points corresponding to HM SFRs.

Figure ~\ref{sanitycheck}b compares the mean surface densities of gas and dust in the SFRs. The picture is less clear here. The HM SFRs display a weak correlation between $\Sigma_{\textrm{dust}}$ and $\Sigma_{\textrm{HI}+\textrm{H}_2}$. The points corresponding to LM SFRs are, on average, located to the right and below, demonstrating the reduced dust abundances in these SFRs, and also the fact that the LM SFRs in our sample are more compact that the HM SFRs.

Figure~\ref{mgas_mdust}a compares the masses of molecular and atomic hydrogen.  There is no correlation between the two states of hydrogen gas. Figure ~\ref{mgas_mdust}b plots the dust mass versus the molecular hydrogen mass (the dust masses were obtained from the IR spectral fitting), and Fig.~\ref{mgas_mdust}c the dust mass versus the atomic hydrogen mass. In the former case, there is a correlation with a Spearman coefficient of 0.75; the points corresponding to LM regions continue the trend for HM regions. In contrast, there is essentially no correlation between the masses of dust and atomic hydrogen. The dust masses for LM and HM regions with the same $M_{\textrm{HI}}$  differ by one to two orders of magnitude.

\begin{figure}[t!]
\includegraphics[width=0.45\textwidth]{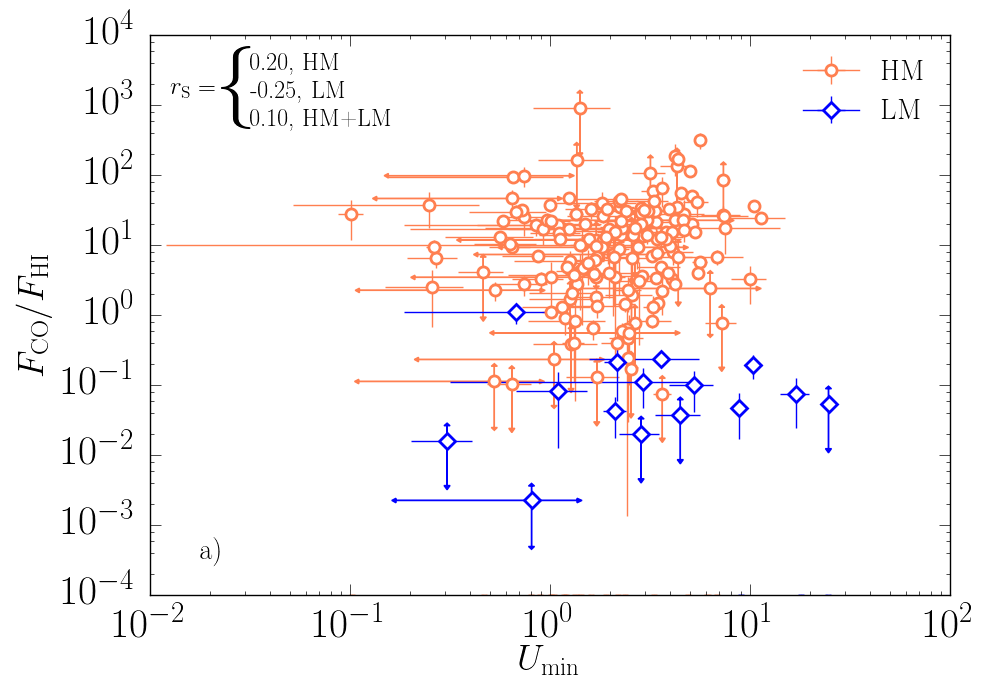}
\includegraphics[width=0.45\textwidth]{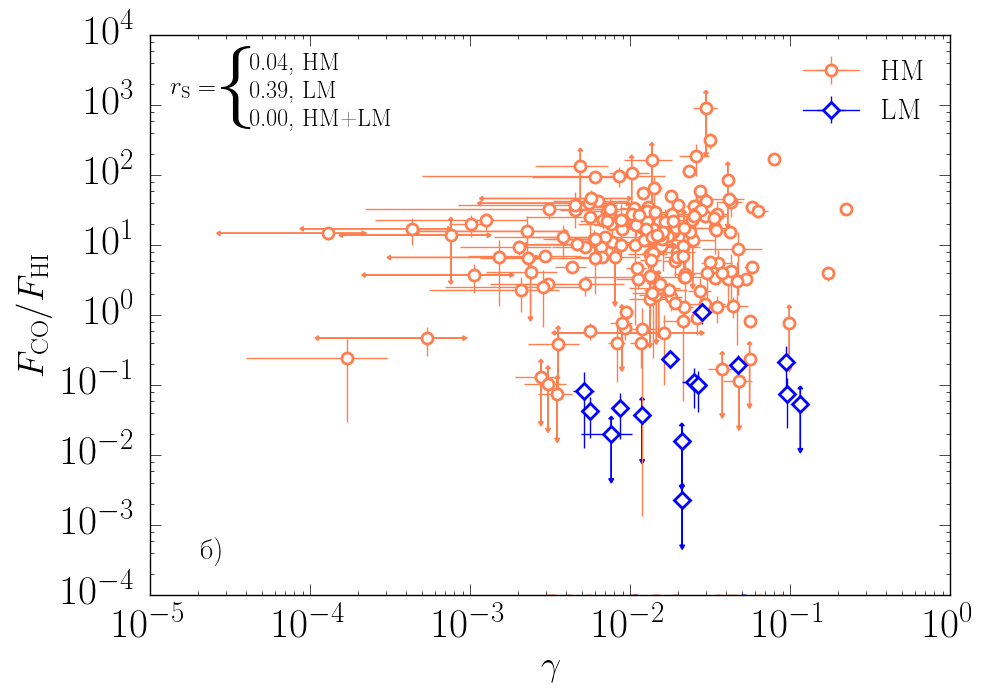}
\includegraphics[width=0.45\textwidth]{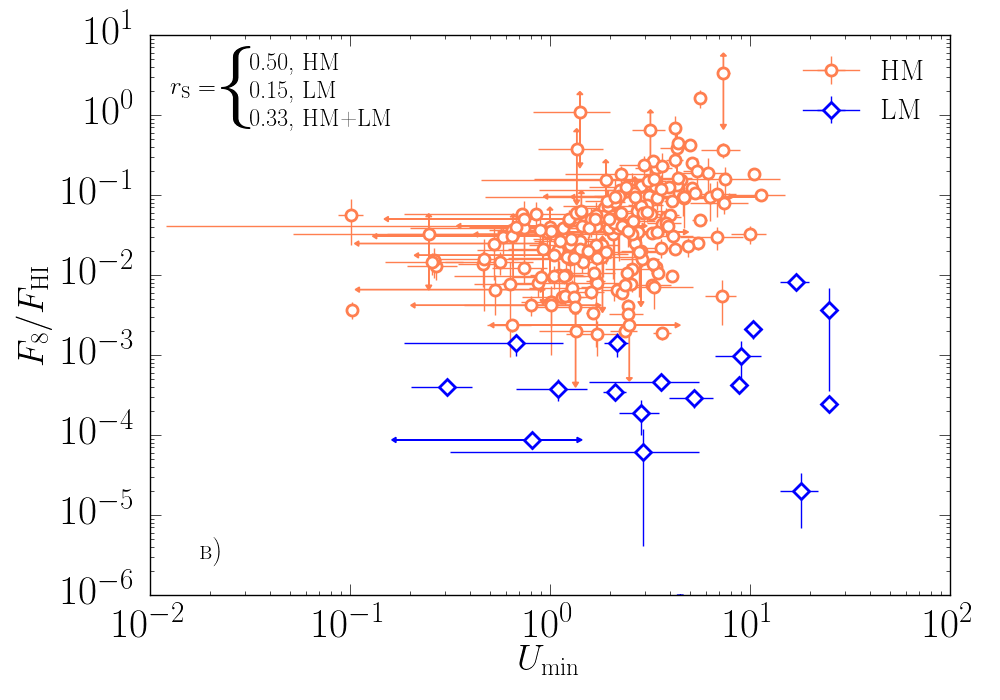}
\includegraphics[width=0.45\textwidth]{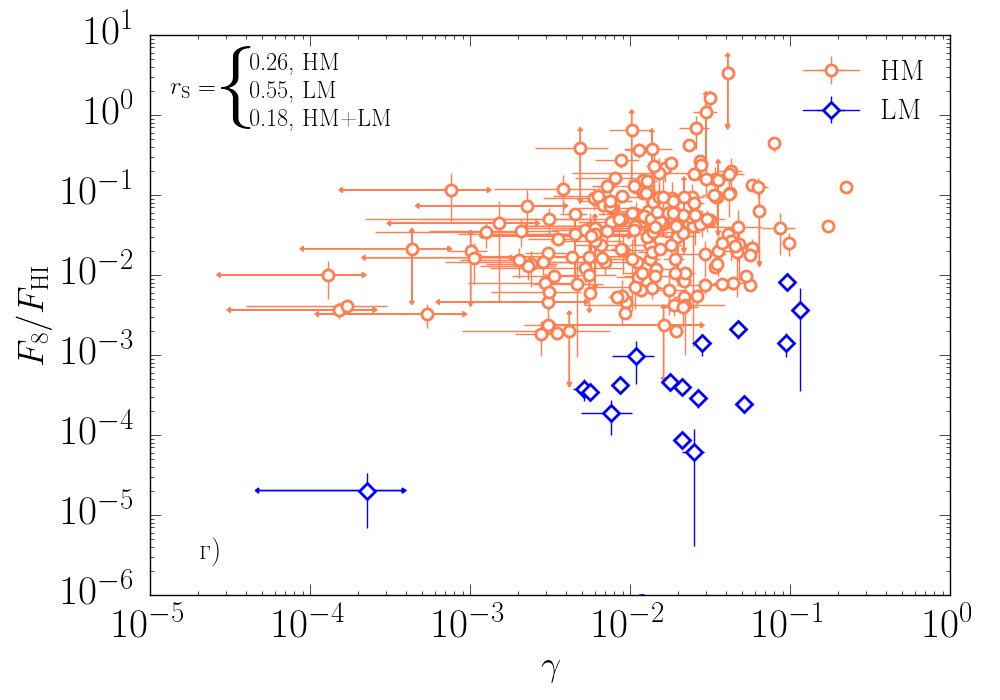}
\includegraphics[width=0.45\textwidth]{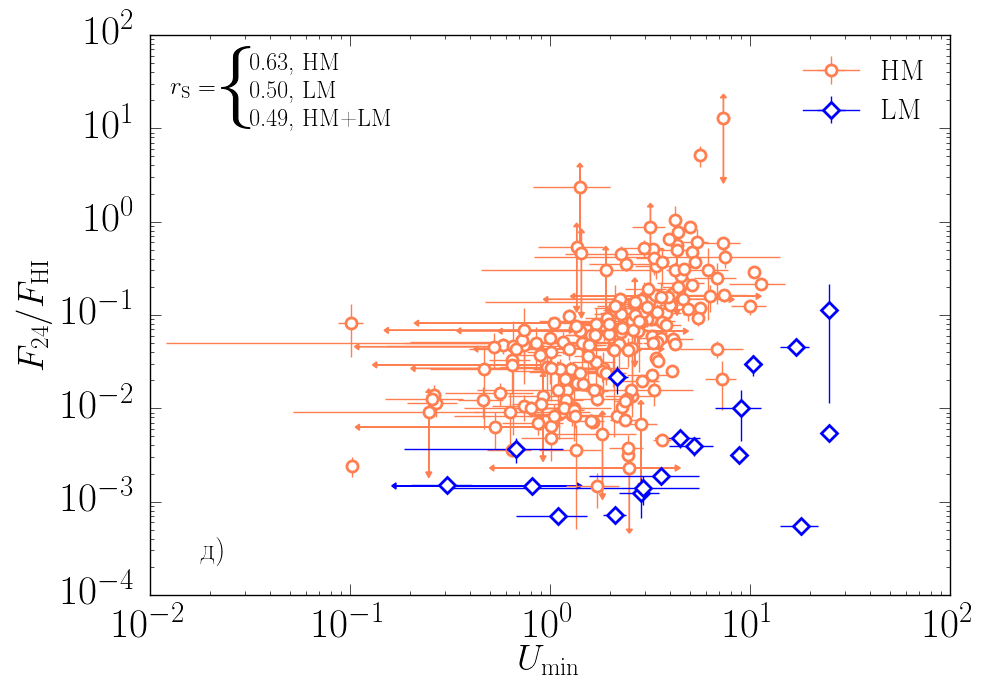}
\includegraphics[width=0.45\textwidth]{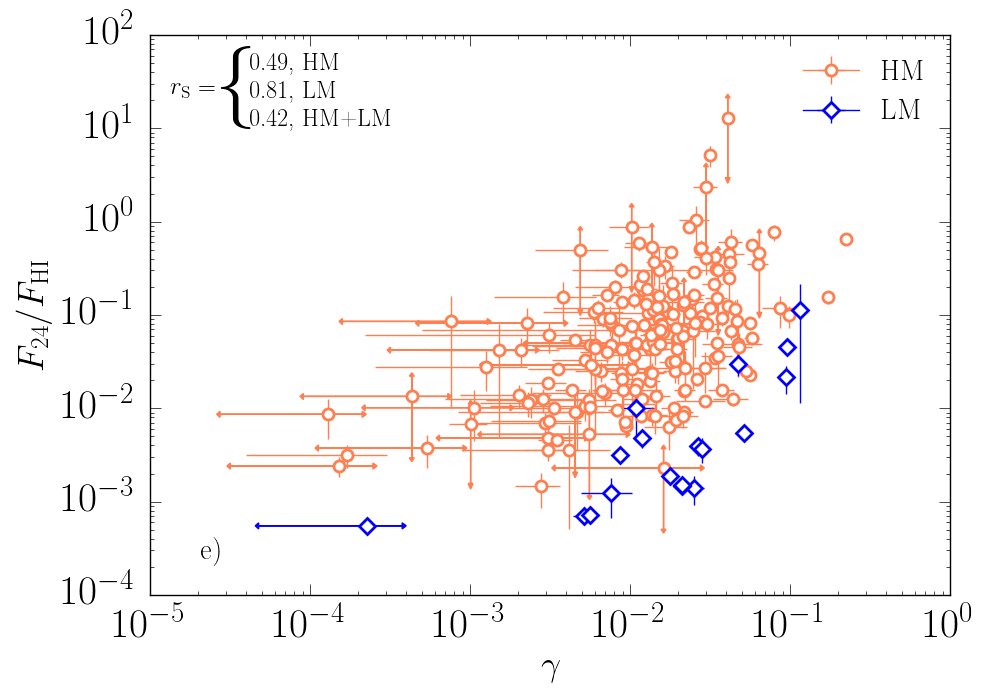}
\includegraphics[width=0.45\textwidth]{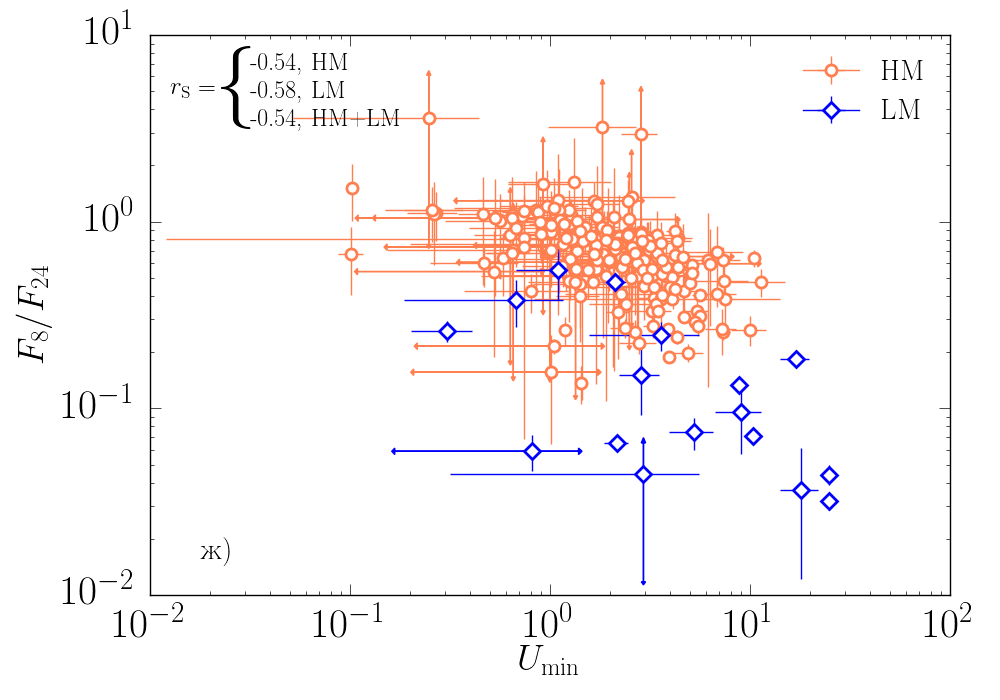}
\includegraphics[width=0.45\textwidth]{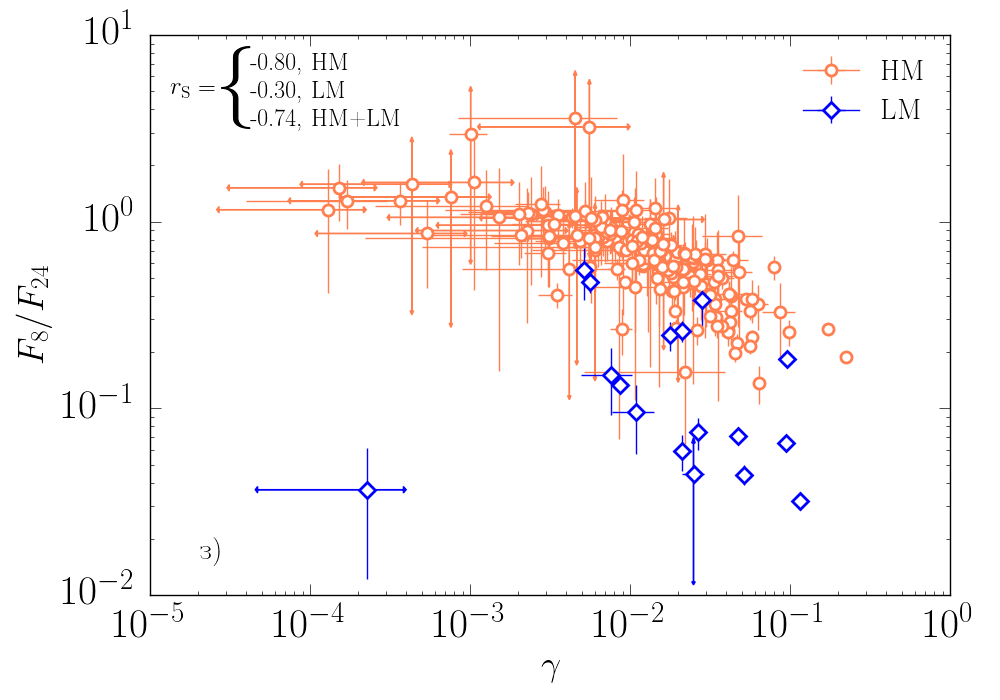}
\caption{Relationships between the parameters $U_{\textrm{min}}$ and $\gamma$ and CO, 8-$\mu$m, and 24-$\mu$m fluxes, normalized to HI fluxes, as well as ratios between 8- and 24-$\mu$m fluxes. \hfill}
\label{ug}
\end{figure}

Results of our IR spectral fitting can also be used to compare various parameters of the SFR emission with the parameters of their radiation fields. Figure~\ref{ug} shows how the CO emission and the ratio of hot dust fluxes at 8 and 24~$\mu$m are related to the parameters $U_{\textrm{min}}$ and $\gamma$. The fluxes are normalized to the SFR fluxes in the HI line.

This normalization requires a special comment. It might be more logical to normalize the fluxes to the derived parameters: the dust mass or H$_2$ mass. However, with the calibration used, the mass of molecular hydrogen is essentially the normalized CO flux corrected for the distance, and there is no sense in comparing this mass with $F_{\textrm{CO}}$. We expected that the dust mass should not depend on the parameters of radiation field, but we found the dust mass to be anticorrelated with $U_{\textrm{min}}$. As was noted in \cite{HolmbergII}, this may be either an artefact of the fitting or a real effect, e.g., representing the destruction of dust in SFRs with strong radiation fields. Accordingly, we selected the HI flux for this normalization, since it does not correlate with the parameters of the radiation fields. A disadvantage of this normalization is that it produces additional scatter in the data. We must also bear in mind the possibility of uncorrected age effects, noted above.

The normalized CO fluxes are essentially uncorrelated with the parameters of the radiation field; the points corresponding to LM SFRs are located below the points for HM SFRs, tracing the dependence of the CO content on metallicity $Z$. The 8 and 24~$\mu$m fluxes grow with both
$U_{\textrm{min}}$ and $\gamma$.On the whole, the correlations between $F_8/F_{\textrm{HI}}$ and the parameters of the radiation field are fairly weak, but the picture becomes more complex when SFRs from both metallicity groups are considered separately. The IR fluxes of the LM SFRs are more highly correlated with $\gamma$, $r_{\textrm{S}}(F_8/F_{\textrm{HI}},\gamma)=0.55$, while for HM SFRs $r_{\textrm{S}}(F_8/F_{\textrm{HI}},\gamma)=0.26$. The Spearman coefficient for the correlation between $F_{24}/F_{\textrm{HI}}$ and $\gamma$ is 0.81 for LM SFRs and 0.49 for HM SFRs. On the contrary, the correlation between $F_{8}/F_{\textrm{HI}}$ and $U_{\textrm{min}}$ is more pronounced for of HM SFRs ($r_{\textrm{S}}$ equal to 0.15 and 0.50 for LM and HM SFRs, respectively). The correlation between the normalized flux $F_{24}/F_{\textrm{HI}}$ and $U_{\textrm{min}}$ is essentially independent of the metallicity ($r_{\textrm{S}}$ $=$ 0.50 and 0.63 for LM and HM SFRs, respectively).

The plots in Fig.~\ref{ug}g and ~\ref{ug}h display the relationships between the parameters of the radiation field and the flux ratio $F_8/F_{24}$. An anticorrelation can be seen for both high and low metallicities: the stronger the radiation field, the lower this ratio. The correlation with $\gamma$ is stronger at high metallicities,($r_{\textrm{S}}(F_8/F_{24},\gamma)={-}0.80$), while the corresponding correlation coefficient at low metallicities is --0.30. The anticorrelation between $U_{\textrm{min}}$ and this flux ratio is approximately independent of metallicity ($r_{\textrm{S}}$ for low and high metallicity equals --0.58 and --0.54, respectively).

\section{DISCUSSION}

We have considered the properties of SFR emission and used them to estimate the physical parameters of extragalactic SFRs. The objects studied are fairly extended, with typical linear sizes of about 500~pc, and with several SFRs having sizes of about 800~pc (the SFRs in our Galaxy have sizes of about 100~pc). The SFRs distinguished in LM galaxies are more compact, with typical sizes of about 200~pc. Accordingly, the SFR masses in LM and HM galaxies are also different. The masses of molecular hydrogen range from $10^5~M_{\odot}$  to $10^8~M_{\odot}$ in HM galaxies, from $10^3~M_{\odot}$  to a few times $\times 10^4~M_{\odot}$ LM galaxies.

The masses of atomic hydrogen are essentially uncorrelated with the masses of molecular hydrogen.
The masses of HI and H$_2$ in HM SFRs are comparable. The masses of molecular gas are substantially lower in LM SFRs, and constitute about $1\%$ of the HI masses. Note that we estimated the H$_2$ masses from the CO fluxes assuming that the conversion factor (``X-factor'') does not depend on metallicity. However, a number of studies show that CO is a poor
indicator of molecular gas in LM galaxies \cite{wolfire03, amorin16, feldman12}. If the X-factor is linearly correlated with the metallicity (approximately in agreement with the observed dependences), we have underestimated $M_{{\textrm{H}}_2}$ in the LM SFRs by about an order of magnitude (which corresponds to the difference inmetallicity between the LM and HM samples). However, even in this case, the difference between the degree of gas molecularization in LM and HM SFRs remains significant (see also Fig.~\ref{obsfluxes3}f).

The dust mass correlates with the mass of molecular hydrogen: in all the SFRs considered, the dust masses constitute about $1\%$ of the molecular gas masses (see Fig. \ref{mgas_mdust}b). Given the size distribution of the dust grains (e.g., \cite{WD01}), this implies that most of the dust mass is contained in large grains. Their temperature is lower than the temperature of small grains, and it is such large grains that are linked to cool, dense molecular gas. However, the dust mass shows no correlation with atomic gas, which means that the observed dust is not associated with the diffuse ISM.

At the same time, the dust mass is correlated with the total mass of molecular and atomic gas, and is in agreement with the trend revealed in observations of galaxies \cite{Draineetal2007}, which can be seen in Fig.~\ref{sanitycheck}. Note the shift toward lower dust masses for the same gas masses in LM SFRs. The decrease in the dustto-gas mass ratio (DGR) in LM galaxies is widely discussed in the literature (\cite{remy2014, feldmann2015,  Draineetal2007}, etc.). The metallicity is probably not the only parameter determining the DGR, and a variety of galactic parameters (star-formation rate, stellar mass, morphological type, radiation field) and evolutionary scenarios for the dust and gas \cite{remy2015}, including dynamical motions \cite{feldmann2015}, should be considered when attempting to understand the DGR. The DGR also varies within a galaxy as a function of the density, temperature, and other ISM parameters \cite{romanduval2014}. However, most of the SFRs we have considered have been spatially unresolved and similar, so that the main factor affecting the DGRs in our sample ismetallicity, ormore precisely, the differences in the evolution of the dust and gas in a LM and HM medium (for example, due to different rates of dustgrain destruction, formation, and growth).

The $U_{\textrm{min}}$ values in the SFRs from our sample vary between approximately 0.1 and 30. Since we consider SFRs dominated by young massive stars, we would expect most of the $U_{\textrm{min}}$ values to be high. Indeed, $U_{\textrm{min}}>1$ in most of the SFRs. Recall that $U_{\textrm{min}}$ is normalized to the Galactic radiation field in the solar vicinity, whereas the mean intensity of the diffuse radiation in other galaxies may be slightly different. It is interesting that the radiation intensity is approximately the same in both LM and HM SFRs,
possibly with a small prevalence of higher $U_{\textrm{min}}$ values in LM SFRs. It is also noteworthy that $U_{\textrm{min}}$ exhibits no correlation with either the SFR masses (e.g., with $M_{\textrm{HI}}$) or their sizes. The values of $\gamma$ cover a wide range, from $10^{-4}$ to ${\sim}0.2$,  also without any visible relationship with metallicity or other global SFR parameters. The observed scatter of the $U_{\textrm{min}}$ and $\gamma$ values may be related  to differences in SFR ages.

As both the PAH and VSG emission is related to re-radiation of the energy of absorbed UV photons, we would expect that the relative fluxes $F_8/F_{\textrm{HI}}$ and $F_{24}/F_{\textrm{HI}}$ should correlate with the parameters of radiation field. However, such a correlation could appear for various reasons. For instance, an increase in the flux with increasing radiation intensity could indicate that brighter emission by dust grains arises in SFRs where the radiation is more intense or harder. On the other hand, a positive correlation between the fluxes and $U_{\textrm{min}}$ and $\gamma$ could be a consequence of the fact that the intensities of both the IR emission and stellar radiation decrease to some degree with the age of the SFR; the IR emission weakens due to the destruction of dust grains, $U_{\textrm{min}}$ and $\gamma$ decrease due to the effects of stellar evolution.
 
Flux ratio $F_8/F_{\textrm{HI}}$ in our sample is appreciably correlated only with $U_{\textrm{min}}$, indicating that the 8~$\mu$m emission is related to the diffuse radiation field formed by all SFR stars in the SFR. This agrees with the conclusions of \cite{Bendoetal2008}, where it was found that PAH emission is most likely associated with the diffusemedium, rather than with the vicinities of young hot stars, whose contribution is characterized by the parameter $\gamma$.
This may be related to the destruction of PAHs near such stars. In other words, PAH destruction may blur the correlation of PAH emission with the fractional volume of SFRs where the intensity of the stellar radiation is higher than $U_{\textrm{min}}$. The correlation between the
normalized fluxes $\gamma$ becomes more pronounced when only LM objects are considered (see Figs. \ref{ug}d, \ref{ug}f). This indicates that many parameters must be taken into account; it is not possible to find a clear correlation with only one parameter, and all of these parameters affect the PAH emission to some degree.

The emission of small hot grains ($F_{24}$) occurs mostly in the vicinities of young stars. Such grains are probably destroyed there less efficiently than PAHs, so that their emission is equally well correlated with $U_{\textrm{min}}$ and $\gamma$.

The contribution of aromatic particles to the total dust mass is estimated using the parameter $q_{\textrm{PAH}}$. We cannot estimate this parameters for the most of SFRs in our sample,  since its values may often be higher or lower than the corresponding limiting values in the sample of synthetic spectra, presented by Draine and Li \cite{2007ApJ...657..810D}. Therefore we can estimate the PAH contribution only from the flux ratio $F_8/F_{24}$. This ratio decreases by more than an order of magnitude with increasing of $U_{\textrm{min}}$ and $\gamma$, possibly reflecting the destruction of small aromatic particles by radiation.

The 8 and 24~$\mu$m fluxes produced by hot grains are well correlated with the CO fluxes, with a greater similarity between the $F_8$ and $F_{\textrm{CO}}$ fluxes. The same result was obtained for the Small Magellanic Cloud in \cite{sandstrom2010}. This may indicate that the histories of the CO and PAH evolution in SFRs are similar. It may be that they are simultaneously and equally efficiently destroyed by the UV emission of young stars. However, the observed picture may also be a consequence of similarity between the conditions for PAH and CO formation. For example, PAHs may form in dense molecular clouds \cite{greenberg2000}, and in SFRs we mainly observe aromatic particles formed {\emph{in situ}}, whereas PAHs formed in stars are destroyed in the ISM by various processes (e.g., shocks from supernova explosions).

A fact that should be specially discussed is the absence of correlation 
between the parameters of stellar radiation field and of the CO (2--1)
emission. Parameter $\gamma$ characterizes radiation fields near
hot massive stars, but CO emission comes from a larger volume, and
therefore the CO/$\gamma$ noncorrelation is expected. However, our
data show the lack of correlation between CO emission and 
the minimum radiation intensity ($U_{\textrm{min}}$), though
the main contribution to the radiation yields the whole 
young stellar population in a SFR. Theoretically, one could
expect the weakening of the CO (2--1) emission with the increase
of $U_{\textrm{min}}$, since stronger radiation more efficiently 
destroys CO molecules, which should affect the observed 
relationships. In addition, the CO (3--2) emission rather than 
the  CO (2--1) emission should be a tracer of dense molecular gas in SFRs,
heated by the cumulative radiation of young stars, since 
the intensity of the latter line decreases with the increase
of temperature. Nevertheless, we could not find any correlation
between the CO emission and radiation field. However, one should
take into account that we consider the CO/HI flux ratio, which
may be influenced by other evolutionary effects.

More detailed analysis is difficult due to at least three
circumstances. First, the most of considered SFRs are fairly
large (about 0.5 kpc) and are not always well separated from
each other. Second, there are no SFRs with intermediate metallicities 
in our sample, and, as a matter of fact, we are dealing 
with two groups of metallicity rather than with a continuous
series. Third, the information about SFR ages will be helpful
in the analysis, but for SFRs from our sample this information 
is absent from the literature. Finally, in the current work we
try to reveal possible correlations between the fluxes of radiation
at different wavelengths, which is a multi-parameter problem.
Several factors can simultaneously influence flux values.
In order to separate the effects of these factors we need a larger
source sample and the data on other parameters (age, hardness 
of radiation etc.). Nevertheless, the presented relationships
can be used for numerical modeling, which will make it possible
to clarify the evolutionary particularities for different components
of gas and dust in extragalactic star-forming regions.

\section{CONCLUSION}

We consider the relationships between the emission of CO molecules, atomic
hydrogen, and dust in the range of wavelengths 8--160~$\mu$m. We used
archival data on approximately 300 star-forming regions in 11 galaxies
with metallicities ($12+\log(\textrm{O/H})$) from 7.54 to 8.71. Data
on the infrared emission of dust are used to determine dust masses
and the parameters of radiation field in the studied SFRs. The following
conclusions can be made based on the obtained results:

1. The content of atomic hydrogen in LM SFRs is higher than in
HM SFRs. The fraction of molecular gas in HM SFRs is comparable
to the fraction of atomic gas, whereas in LM SFRs the fraction
of molecular gas is an order of magnitude or more (depending
on the method of estimating the mass of molecular gas) lower
than that of atomic gas and may be less than $10\%$.

2. The masses of dust in all the considered SFRs are about $1\%$
of the masses of atomic gas. The mass of dust also correlates
with the total mass of atomic and molecular gas, being
in agreement with the trend revealed for galaxies as a whole. 
The masses and surface densities of dust in LM SFRs are significantly
lower than in HM SFRs with the same total masses of gas.

3. Fluxes of radiation in the NIR (8~$\mu$m), MIR (24~$\mu$m) and
FIR (70, 100 and 160~$\mu$m) ranges are well correlated between each
other, pointing on a common origin of radiation in these ranges: 
probably, all this radiation arises in hot dust in the vicinities
of young stars. When FIR fluxes in HM and LM SFRs are the same,
the 8~$\mu$m emission in the latter SFRs is weaker, confirming
the known correlation between the metallicity and the abundance
of PAHs, which are considered to be the main source of emission
at 8~$\mu$m. Parameters of emission at 24~$\mu$m do not depend
on metallicity.

4. CO fluxes virtually do not correlate with the parameters of
radiation field in SFRs. IR fluxes at 8 and 24~$\mu$m increase
with increasing both the minimum level of UV intensity in SFRs
and the fraction of dust illuminated by radiation 
that has intensity above this level. This correlation may represent
both more intensive emission of hot dust grains in medium with 
higher number of exciting photons and an evolutionary decrease
of the flux going simultaneously with the decrease of radiation
intensity. Flux ratio $F_8/F_{24}$ demonstrates an anticorrelation
with parameters that describe the intensities of radiation in SFRs. 
If this ratio represents the content of aromatic grains in SFRs,
this anticorrelarion may be an evidence for their destruction in
intensive radiation fields.

The work was partly supported by RFBR grants $No~16-32-00237$ and
13-02-00466, by the grants of the President of the Russian Federation 
MK-4536.2015.2 and NSh-9576.2016.2, by the program 211 of 
the government of the Russian Federation, agreement $No~02.A03.21.0006$, 
by the Ministry of Education and Science of the Russian Federation 
(the project part of the government mandate $No~3.1781.2014/K$).
The work is based on the observations of Spitzer Space Telescope,
developed in the Jet Propulsion Laboratory of the California 
Institute of Technology under a contract with NASA, of Herschel
Space Observatory, built by ESA under NASA support, of the VLA 
telescope of NRAO, which is a facility of the National Science 
Foundation, and of the 30-m IRAM radio telescope. The astronomical 
database HYPERLEDA (http://leda.univ-lyon1.fr) is used in the work.


\clearpage
\begin{landscape}
\begin{table}[h]\tiny
\caption{Photometric results for the galaxy Holmberg II}
\label{phot_HoII}
\begin{tabular}{|l|c|c|c|c|c|c|c|c|c|c|c|c|c|c|c|c|c|c|c|c|c|c|c|}
\hline
 n & $\alpha$ & $\delta$ & $R$	&	$F_{3.6}\pm\Delta F_{3.6}$	&	$F_{4.5}\pm\Delta F_{4.5}$	&	$F_{5.8}\pm\Delta F_{5.8}$	&	$F_{8}\pm\Delta F_{8}$	&	$F_{8}^{\rm afe}\pm\Delta F_{8}^{\rm afe}$	&	$F_{24}\pm\Delta F_{24}$	&	$F_{24}^{\rm ns}\pm\Delta F_{24}^{\rm ns}$	&	$F_{70}\pm\Delta F_{70}$ & $F_{100}\pm\Delta F_{100}$ & $F_{160}\pm\Delta F_{160}$ &	$F_{\mathrm{CO}}\pm\Delta F_{\mathrm{CO}}$	&	$F_{\mathrm{HI}}\pm\Delta F_{\mathrm{HI}}$\\
 &  &  &	& mJy & mJy & mJy & mJy & mJy &	mJy & mJy & mJy & mJy & mJy & Jy $\cdot$ km/s  & Jy $\cdot$ km/s\\
\hline
1	&	 124.70748	&	70.746922	&	12" 	&	0.19	 $\pm$ 	0.02	&	0.22	 $\pm$ 	0.02	&	0.44	 $\pm$ 	0.03	&	0.64	 $\pm$ 	0.05	&	0.41	 $\pm$ 	0.04	&	9.32	 $\pm$ 	0.52	&	9.25	 $\pm$ 	0.52	&	74.51	 $\pm$ 	3.31	&	70.92	 $\pm$ 	4.36	&	34.32	 $\pm$ 	2.50	&	0.00	 $\pm$ 	0.03	&	1.69	 $\pm$ 	0.06	\\
2	&	 124.72882	&	70.719817	&	12" 	&	0.21	 $\pm$ 	0.06	&	0.14	 $\pm$ 	0.05	&	0.15	 $\pm$ 	0.04	&	0.06	 $\pm$ 	0.01	&	0.00	 $\pm$ 	0.03	&	0.98	 $\pm$ 	0.23	&	0.90	 $\pm$ 	0.23	&	3.93	 $\pm$ 	4.62	&	5.21	 $\pm$ 	7.64	&	6.14	 $\pm$ 	6.61	&	0.11	 $\pm$ 	0.02	&	0.11	 $\pm$ 	0.08	\\
3	&	 124.72882	&	70.716097	&	12" 	&	0.25	 $\pm$ 	0.05	&	0.20	 $\pm$ 	0.03	&	0.21	 $\pm$ 	0.04	&	0.13	 $\pm$ 	0.01	&	0.00	 $\pm$ 	0.02	&	2.04	 $\pm$ 	0.16	&	1.95	 $\pm$ 	0.16	&	26.30	 $\pm$ 	3.88	&	38.92	 $\pm$ 	6.38	&	34.51	 $\pm$ 	4.47	&	0.02	 $\pm$ 	0.03	&	0.40	 $\pm$ 	0.08	\\
4	&	 124.73447	&	70.710631	&	12" 	&	0.07	 $\pm$ 	0.05	&	0.05	 $\pm$ 	0.04	&	0.14	 $\pm$ 	0.04	&	0.10	 $\pm$ 	0.01	&	0.04	 $\pm$ 	0.03	&	1.12	 $\pm$ 	0.16	&	1.09	 $\pm$ 	0.16	&	28.48	 $\pm$ 	4.14	&	41.16	 $\pm$ 	4.47	&	20.00	 $\pm$ 	5.38	&	0.00	 $\pm$ 	0.03	&	1.98	 $\pm$ 	0.26	\\
5	&	 124.75129	&	70.717389	&	12" 	&	0.13	 $\pm$ 	0.04	&	0.09	 $\pm$ 	0.03	&	0.12	 $\pm$ 	0.04	&	0.20	 $\pm$ 	0.02	&	0.11	 $\pm$ 	0.03	&	0.78	 $\pm$ 	0.20	&	0.73	 $\pm$ 	0.20	&	20.69	 $\pm$ 	2.64	&	32.01	 $\pm$ 	4.53	&	35.03	 $\pm$ 	3.78	&	0.01	 $\pm$ 	0.03	&	0.59	 $\pm$ 	0.22	\\
6	&	 124.76345	&	70.717172	&	12" 	&	0.14	 $\pm$ 	0.04	&	0.09	 $\pm$ 	0.03	&	0.15	 $\pm$ 	0.04	&	0.12	 $\pm$ 	0.03	&	0.03	 $\pm$ 	0.03	&	0.82	 $\pm$ 	0.12	&	0.77	 $\pm$ 	0.12	&	10.65	 $\pm$ 	3.40	&	18.63	 $\pm$ 	4.74	&	16.94	 $\pm$ 	5.22	&	0.06	 $\pm$ 	0.03	&	0.55	 $\pm$ 	0.17	\\
7	&	 124.84857	&	70.716375	&	12" 	&	0.23	 $\pm$ 	0.12	&	0.16	 $\pm$ 	0.07	&	0.18	 $\pm$ 	0.08	&	0.06	 $\pm$ 	0.02	&	-0.06	 $\pm$ 	0.05	&	0.81	 $\pm$ 	0.80	&	0.73	 $\pm$ 	0.81	&	3.74	 $\pm$ 	8.82	&	6.73	 $\pm$ 	12.05	&	6.50	 $\pm$ 	5.92	&	0.00	 $\pm$ 	0.04	&	0.51	 $\pm$ 	0.23	\\
8	&	 124.84639	&	70.698353	&	12" 	&	0.13	 $\pm$ 	0.16	&	0.07	 $\pm$ 	0.10	&	0.09	 $\pm$ 	0.10	&	0.01	 $\pm$ 	0.01	&	0.00	 $\pm$ 	0.07	&	0.09	 $\pm$ 	0.21	&	0.04	 $\pm$ 	0.22	&	0.60	 $\pm$ 	5.23	&	0.00	 $\pm$ 	3.86	&	0.00	 $\pm$ 	2.46	&	0.00	 $\pm$ 	0.03	&	0.00	 $\pm$ 	0.14	\\
9	&	 124.86434	&	70.699719	&	12" 	&	0.06	 $\pm$ 	0.13	&	0.03	 $\pm$ 	0.10	&	0.09	 $\pm$ 	0.08	&	0.12	 $\pm$ 	0.02	&	0.07	 $\pm$ 	0.06	&	2.37	 $\pm$ 	0.23	&	2.35	 $\pm$ 	0.23	&	33.36	 $\pm$ 	4.23	&	15.89	 $\pm$ 	4.14	&	11.48	 $\pm$ 	2.51	&	0.04	 $\pm$ 	0.03	&	0.40	 $\pm$ 	0.16	\\
10	&	 124.8654	&	70.705728	&	12" 	&	0.57	 $\pm$ 	0.28	&	0.36	 $\pm$ 	0.18	&	0.32	 $\pm$ 	0.16	&	0.14	 $\pm$ 	0.04	&	0.00	 $\pm$ 	0.13	&	1.13	 $\pm$ 	0.43	&	0.94	 $\pm$ 	0.44	&	25.51	 $\pm$ 	6.29	&	21.90	 $\pm$ 	6.82	&	6.31	 $\pm$ 	4.50	&	0.00	 $\pm$ 	0.03	&	0.00	 $\pm$ 	0.16	\\
11	&	 124.87076	&	70.716944	&	12" 	&	0.28	 $\pm$ 	0.05	&	0.35	 $\pm$ 	0.04	&	0.87	 $\pm$ 	0.05	&	1.16	 $\pm$ 	0.07	&	0.75	 $\pm$ 	0.07	&	23.41	 $\pm$ 	1.55	&	23.31	 $\pm$ 	1.55	&	91.40	 $\pm$ 	10.27	&	65.29	 $\pm$ 	14.20	&	29.30	 $\pm$ 	7.20	&	0.01	 $\pm$ 	0.03	&	0.21	 $\pm$ 	0.19	\\
12	&	 124.80287	&	70.718733	&	12" 	&	0.77	 $\pm$ 	0.42	&	1.89	 $\pm$ 	0.20	&	4.36	 $\pm$ 	0.31	&	6.03	 $\pm$ 	0.24	&	4.00	 $\pm$ 	0.30	&	22.02	 $\pm$ 	1.35	&	21.76	 $\pm$ 	1.35	&	139.42	 $\pm$ 	8.75	&	130.92	 $\pm$ 	8.73	&	85.69	 $\pm$ 	6.37	&	0.04	 $\pm$ 	0.02	&	0.48	 $\pm$ 	0.06	\\

\hline
\end{tabular}
\end{table}

\begin{table}[h!]
\caption{Dust and gas parameters in the galaxy Holmberg II}
\label{galdata}
\begin{center}
\begin{tabular}{|l|c|c|c|c|c|c|}
\hline
$n$ & 12 + log(O/H)  & $q_{\textrm{PAH}}\pm\Delta
q_{\textrm{PAH}}$, $\%$& \multicolumn{1}{c|}{$U_{\textrm{min}}
\pm \Delta U_{\textrm{min}}$}    &
\multicolumn{1}{c|}{$\gamma\pm\Delta \gamma  \times 10^{-3}$} &
\multicolumn{1}{c|}{$M_{\textrm{dust}}\pm\Delta
M_{\textrm{dust}}$, $M_{\odot}$}&
\multicolumn{1}{c}{$\Omega\pm\Delta \Omega$}    \\
\hline
1	&	7.72	&	0.41  	$\pm$	0.00  	&	24.69  	$\pm$	0.13  	&	51.34  	$\pm$	0.42  	&	30.61  	$\pm$	0.14  	&	30.24  	$\pm$	0.17  	\\
3	&	7.72	&	0.45  	$\pm$	0.00  	&	4.48  	$\pm$	1.13  	&	11.91  	$\pm$	1.31  	&	64.51  	$\pm$	19.01  	&	23.25  	$\pm$	2.12  	\\
4	&	7.72	&	0.42  	$\pm$	0.00  	&	17.92  	$\pm$	3.87  	&	0.23  	$\pm$	1.22  	&	21.05  	$\pm$	4.48  	&	4.84  	$\pm$	1.65  	\\
5	&	7.72	&	0.72  	$\pm$	0.10  	&	2.86  	$\pm$	0.65  	&	7.64  	$\pm$	2.72  	&	106.10  $\pm$	21.16  	&	17.38  	$\pm$	2.59  	\\
6	&	7.72	&	0.46  	$\pm$	0.00  	&	2.93  	$\pm$	2.62  	&	24.94  	$\pm$	3.93  	&	49.94  	$\pm$	34.45  	&	19.40  	$\pm$	2.53  	\\
11	&	7.72	&	0.47  	$\pm$	0.00  	&	24.69  	$\pm$	0.12  	&	116.10  $\pm$	0.69  	&	27.06  	$\pm$	0.13  	&	53.92  	$\pm$	0.43  	\\
12	&	7.72	&	2.71  	$\pm$	0.12  	&	16.96  	$\pm$	2.79  	&	95.68  	$\pm$	8.80  	&	83.68  	$\pm$	10.87  	&	301.80  $\pm$	23.47  	\\ \hline
\end{tabular}
\end{center}
\end{table}

\clearpage
\end{landscape}

\end{document}